%% file: main.tex
\UseRawInputEncoding
 \documentclass[sigconf,screen,nocopyrightspace]{acmart}
\AtBeginDocument{%
  \providecommand\BibTeX{{%
    \normalfont B\kern-0.5em{\scshape i\kern-0.25em b}\kern-0.8em\TeX}}}

\usepackage{listings}
\usepackage{multirow}

\usepackage{mathtools}
\DeclarePairedDelimiter{\ceil}{\lceil}{\rceil}

\usepackage{threeparttable}

\usepackage{xcolor}

\usepackage{subcaption,graphicx}

\usepackage{pifont}

\newcommand{\new}[1]{\textcolor{black}{#1}}

\begin{document}

\title{Dagger: Accelerating RPCs in Cloud Microservices Through Tightly-Coupled Reconfigurable NICs}

\author{Nikita Lazarev}
\affiliation{%
  \institution{Cornell University}
  \city{Ithaca}
  \state{New York}
  \country{USA}
  \postcode{14850}
}
\email{nl524@cornell.edu}

\author{Shaojie Xiang}
\affiliation{%
  \institution{Cornell University}
  \city{Ithaca}
  \state{New York}
  \country{USA}
  \postcode{14850}
}
\email{sx233@cornell.edu}

\author{Neil Adit}
\affiliation{%
  \institution{Cornell University}
  \city{Ithaca}
  \state{New York}
  \country{USA}
  \postcode{14850}
}
\email{na469@cornell.edu}

\author{Zhiru Zhang}
\affiliation{%
  \institution{Cornell University}
  \city{Ithaca}
  \state{New York}
  \country{USA}
  \postcode{14850}
}
\email{zhiruz@cornell.edu}

\author{Christina Delimitrou}
\affiliation{%
  \institution{Cornell University}
  \city{Ithaca}
  \state{New York}
  \country{USA}
  \postcode{14850}
}
\email{delimitrou@cornell.edu}



\begin{abstract}
The ongoing shift of cloud services from monolithic designs to microservices creates high demand for efficient and high performance datacenter networking stacks, optimized for fine-grained workloads. Commodity networking systems based on software stacks and peripheral NICs introduce high overheads when it comes to delivering small messages. 

We present \textit{Dagger}, a hardware acceleration fabric for cloud RPCs based on FPGAs, where the accelerator is closely-coupled with the host processor over a configurable memory interconnect. The three key design principle of Dagger are: (1) offloading the entire RPC stack to an FPGA-based NIC, (2) leveraging memory interconnects instead of PCIe buses as the interface with the host CPU, and (3) making the acceleration fabric reconfigurable, so it can accommodate the diverse needs of microservices. We show that the combination of these principles significantly improves the efficiency and performance of cloud RPC systems \new{while preserving their generality}. Dagger achieves $1.3-3.8\times$ higher per-core RPC throughput compared to both highly-optimized software stacks, and systems using specialized RDMA adapters. It also scales up to 84 Mrps with 8 threads on 4 CPU cores, while maintaining state-of-the-art $\mu$s-scale tail latency. \new{We also demonstrate that large third-party applications, like memcached and MICA KVS, can be easily ported on Dagger with minimal changes to their codebase, bringing their median and tail KVS access latency down to $2.8 - 3.5$ us and $5.4 - 7.8$ us, respectively. Finally, we show that Dagger is beneficial for multi-tier end-to-end microservices with different threading models by evaluating it using an 8-tier application implementing a flight check-in service. }
\end{abstract}



\maketitle

\input{Introduction}

\input{Related_Work}

\input{Network_Characterization}

\input{Dagger_Design}

\input{Evaluation}

\input{Discussions}

\input{Conclusions}


\bibliographystyle{ACM-Reference-Format}
\bibliography{sample-base}


\end{document}

%% file: Introduction.tex
\section{Introduction}
\label{sec:intro}

Modern cloud applications are increasingly turning to the microservices programming model to improve their agility, elasticity, and modularity~\cite{uTune,DeathStarBench,Cockroft16,Sriraman2018SA,Cockroft15,twitter_decomposing,Gan18,Gan19b,Lazarev20,Chen19,xcontainers,Zhang19,Gan21,Zhang21,Delimitrou13,Delimitrou14,Gan18c,Delimitrou15,Delimitrou16,Delimitrou17}. Microservices break complex monolithic applications, which implement the entire functionality as a single service, into many fine-grained and loosely-coupled tiers. 
This improves design modularity, error isolation, and facilitates development and deployment. However, since microservices communicate with each other over the network, they also introduce significant communication overheads~\cite{DeathStarBench,OptimusPrime}. Given that individual microservices typically involve a small amount of computation, networking ends up being a large fraction of their overall latency~\cite{DeathStarBench, Sriraman2018SA}. Furthermore, since microservices depend on each other, performance unpredictability due to network congestion can propagate across dependent tiers and degrade the end-to-end performance~\cite{Gan19b,causeinfer,microscope,explainit,sosp03blackbox}. 

Microservices typically communicate with each other over Remote Procedure Calls (RPC)~\cite{Thrift,gRPC,finagle}. Unfortunately, existing RPC frameworks were not designed specifically for microservices, whose network requirements and traffic characteristics differ from traditional cloud applications, and therefore introduce significant overheads to their performance. The strict latency requirements, fine-grained requests, wide diversity, and frequent design cadence of microservices put a lot of pressure on the network system, and makes rethinking networking with microservices in mind a pressing need. Most of the existing commercial RPC frameworks are implemented on top of commodity OS networking, such as Linux TCP/IP. While this ensures generality, such systems suffer from considerable overheads across all levels of the system stack~\cite{IX,OptimusPrime}. These overheads accumulate over deep microservice call paths, and result in end-to-end QoS violations. While this affects all microservices, it is especially challenging for interactive tiers, which optimize for low tail latency, instead of average throughput.

The past decade has seen increased interest both from academia and industry for lower latency and higher throughput networking systems. One line of work focuses on optimizing transport protocols~\cite{HOMA, mTCP, DCTCP,pFABRIC}, while another moves networking to user space~\cite{IX, DBLP:conf/nsdi/KaliaKA19, Seastar,mTCP}, or offloads it to specialized adapters~\cite{FARM, FASST,Ming2012UserspaceRO, AccelTCP,HotNet}. Network programmability has also gained traction through the use of SmartNICs~\cite{P_Costa, E3,msft} to tune the network configuration to the performance requirements of target applications. Despite the performance and efficiency benefits of these approaches, they are limited in the type of interfaces they use between the host CPU and the NIC. Almost all commercially available NICs are viewed by the processor as PCIe-connected peripheral devices. Unfortunately, PCIe interconnects require multiple bus transactions, memory synchronizations, and expensive MMIO requests for every request to the NIC~\cite{rdma_1, Cambridge, stanford_mmio, NetDIMM}. As a result, the per-packet overhead in these optimized systems is still high; this is especially noticeable for fine-grained workloads with deep call paths, like microservices. 

This paper presents \textit{Dagger}, an FPGA-based reconfigurable RPC stack integrated into the processor's memory subsystem over a NUMA interconnect. Integrated and near-memory NICs have already shown promise in reducing the overheads of PCIe, and improving networking efficiency~\cite{NetDIMM, soNUMA, NeBuLa}. However, prior integrated NICs are based on ASICs that lack reconfigurability, and require taping out custom chips, which is expensive and time consuming for frequently-changing networking configurations at datacenter scale. Widely-used RPC stacks for microservices, such as Thrift RPC~\cite{Thrift}, gRPC~\cite{gRPC}, offer a rich variety of transport options, (de)serialization methods, and threading models. Hardware-based RPC stacks can only be practical in the context of microservices if they allow the same flexibility. To this end, we propose an \textit{integrated} and \textit{reconfigurable} FPGA-accelerated networking fabric, capable of supporting realistic and end-to-end RPC frameworks. 

Dagger is based on three key design principles: (1) The NIC implements the entire RPC stack in hardware, while the software is only responsible for providing the RPC API. This way, we remove CPU-related overheads from the critical path of RPC flows, and free more CPU resources for the high concurrency of microservices. (2) Dagger leverages memory interconnects to communicate with the processor. We show that in contrast to PCIe protocols that were initially designed for the Producer-Consumer dataflow pattern, memory interconnects offer a better communication model that is especially beneficial for transferring ready-to-use RPC objects. We also argue that integrating NICs via memory interconnects is more practical than previously-proposed methods of closely-coupling NICs with CPUs, since processors today come with exposed memory busses, with the next generation of server-class CPUs already offering dedicated peripheral memory interconnects~\cite{CXL}. (3) Finally, Dagger is based on an FPGA, so its design is fully programmable. This allows it to adjust to the performance and resource requirements of a given microservice. 

We build Dagger on an Intel Broadwell CPU/FPGA hybrid architecture, similar to those available in public clouds, such as Intel HARP. We show that offloading the entire RPC stack to hardware enables better CPU efficiency, which results in higher per-core RPC throughput and lower request latency. In addition, we demonstrate the benefits of closely-coupling hardware RPC stacks with applications through memory interconnects. Dagger improves the per-core RPC throughput up to 1.3-3.8$\times$ compared to prior work, based on both optimized software RPC frameworks~\cite{DBLP:conf/nsdi/KaliaKA19} and specialized hardware adapters~\cite{FASST}. Dagger reaches 12.4 - 16.5 Mrps of per core throughput, and it scales up to 42 Mrps with only 4 physical threads on two CPU cores, while achieving state-of-the-art $\mu$s-scale end-to-end latency. 

In addition, we show that Dagger can be easily integrated into existing datacenter applications with minor changes to the codebase. \new{Our experiments with memcached and MICA KVS using Dagger as the communication layer show that it achieves median and $99^{th}$ percentile tail latency of 3.2 and 7.8 us respectively for memcached and 3.5 us and 5.7 us for MICA, while also achieving a throughput of 5.2 Mrps~\cite{memcached} on a single core. This result is $11.4\times$ lower than the latency of memcached over a native transport based on the Linux kernel networking, and $4.4 - 5.2\times$ lower than of MICA over a highly-optimized, DPDK-based user space networking stack. Finally, we demonstrate that Dagger can accommodate multi-tier microservice applications with diverse requirements and threading models by porting an 8-tier Flight Registration service on top of Dagger, and showing significant performance benefits compared to native execution. }

%% file: Related_Work.tex
\section{Related Work}
\label{sec:related}

Designing low-latency networking systems including optimizations for small requests is not a new problem. Both industry and academia have contributed various proposals to this end. In this section, we briefly review prior work on networking acceleration, and discuss how it resembles and differs from our proposal. We classify related work into three categories: (1) software solutions, (2) systems leveraging specialized commercial hardware adapters, and (3) proposals of new hardware architectures for efficient networking.

\vspace{0.05in}
\noindent{\bf{Software-based solutions: }}
At the software level, most research has focused on transport protocol optimizations for low latency and/or high throughput~\cite{DCTCP, mTCP, HOMA, pFABRIC, RestructuringCC}. This includes optimizing the congestion control mechanisms, flow scheduling, connection management, etc. In addition to transport optimizations, some proposals also suggest moving networking from the kernel to user space by leveraging, for example, DPDK~\cite{DPDK, HOMA, mTCP, IX, TAS} or raw NIC driver APIs~\cite{DBLP:conf/nsdi/KaliaKA19}. Although these proposals demonstrate their efficiency in improving the performance of datacenter networks, they are still subject to system overheads due to their software-only and CPU-based implementation.

\vspace{0.05in}
\noindent{\bf{Specialized commercial adapters:}} As an alternative to software/ algorithmic-only optimizations, another line of work proposed to leverage RDMA hardware to offload network processing to specialized adapters, and use the remote memory abstraction to implement higher-level communication primitives, such as RPCs~\cite{FARM, FASST, DaRPC}. This approach improves CPU efficiency, and as a consequence also incurs lower latency and higher throughput. Despite this, there are two main issues with existing RDMA-related work. First, prior work does not implement a fully-offloaded RPC protocol; commodity RDMA adapters only offload the networking part, i.e., up to the transport layer and RDMA protocols, keeping the execution of RPC layers on the host CPU. Second, all RDMA NICs are seen as peripheral devices from the perspective of the host CPUs, and are interconnected with the latter over PCIe busses that have been shown to be inefficient, especially when the metric of interest is latency and network packets are small~\cite{rdma_1, stanford_mmio, Cambridge}. Therefore, bringing NICs closer to CPUs and/or memory is required to enable efficient and fast communication in datacenters.

\begin{figure}[h]
  \centering
  \includegraphics[width=0.46\textwidth]{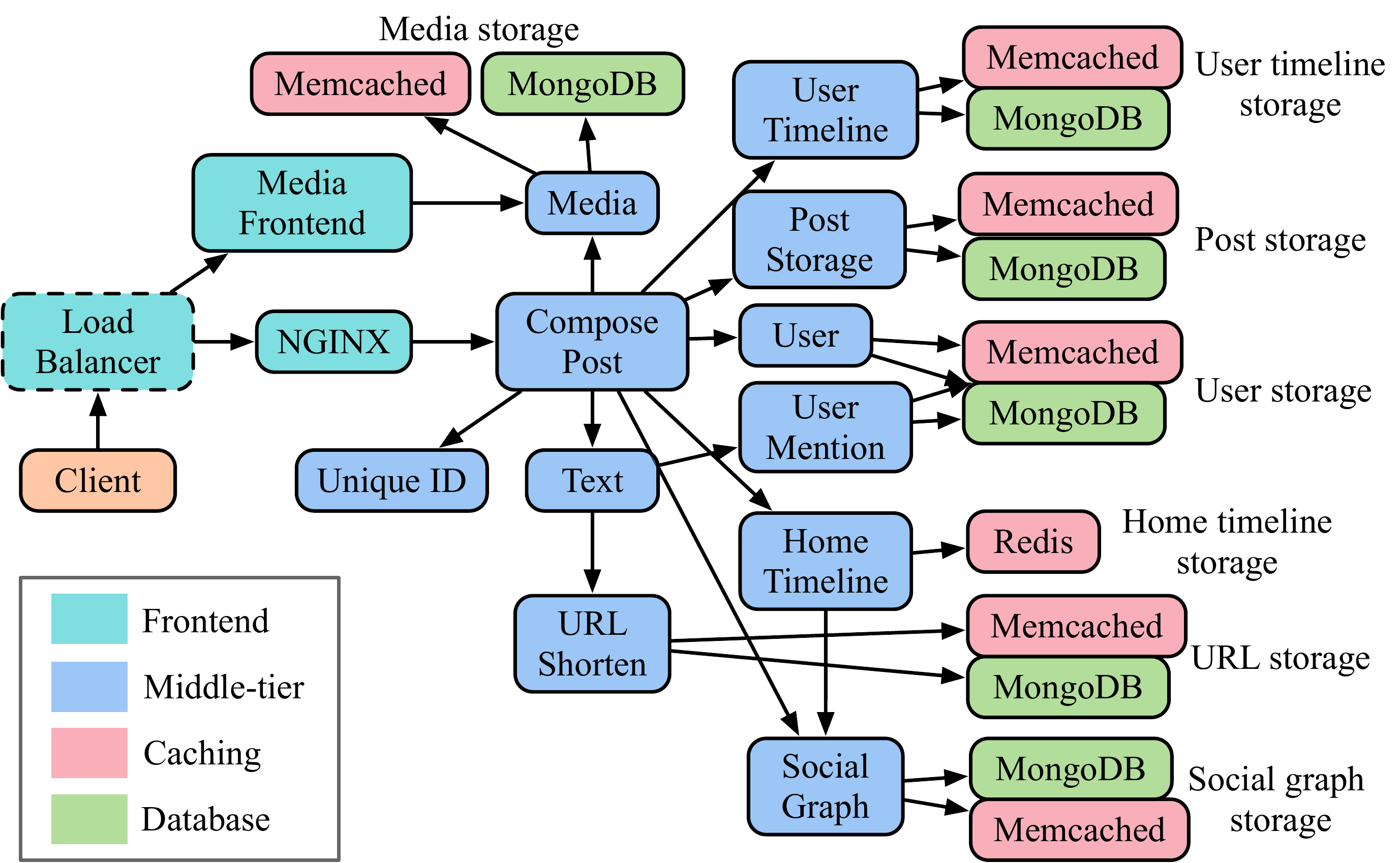}
  \caption{\new{Social Network microservice architecture~\cite{DeathStarBench} -- \normalfont{client requests first reach a front-end load balancer, which evenly distributes them across the $N$ webserver instances. Then, depending on the type of user request, mid-tiers will be invoked to create a post, read a user's timeline, follow/unfollow users, or receive recommendations on new users to follow. At the right-most of the figure, the requests reach the back-end databases, implemented both with in-memory caching tiers (memcached/Redis), and persistent databases (MongoDB). }}}
  \label{fig:social_network}
\end{figure}

\vspace{-5pt}

\begin{figure}[h]
  \centering
  \includegraphics[width=0.46\textwidth]{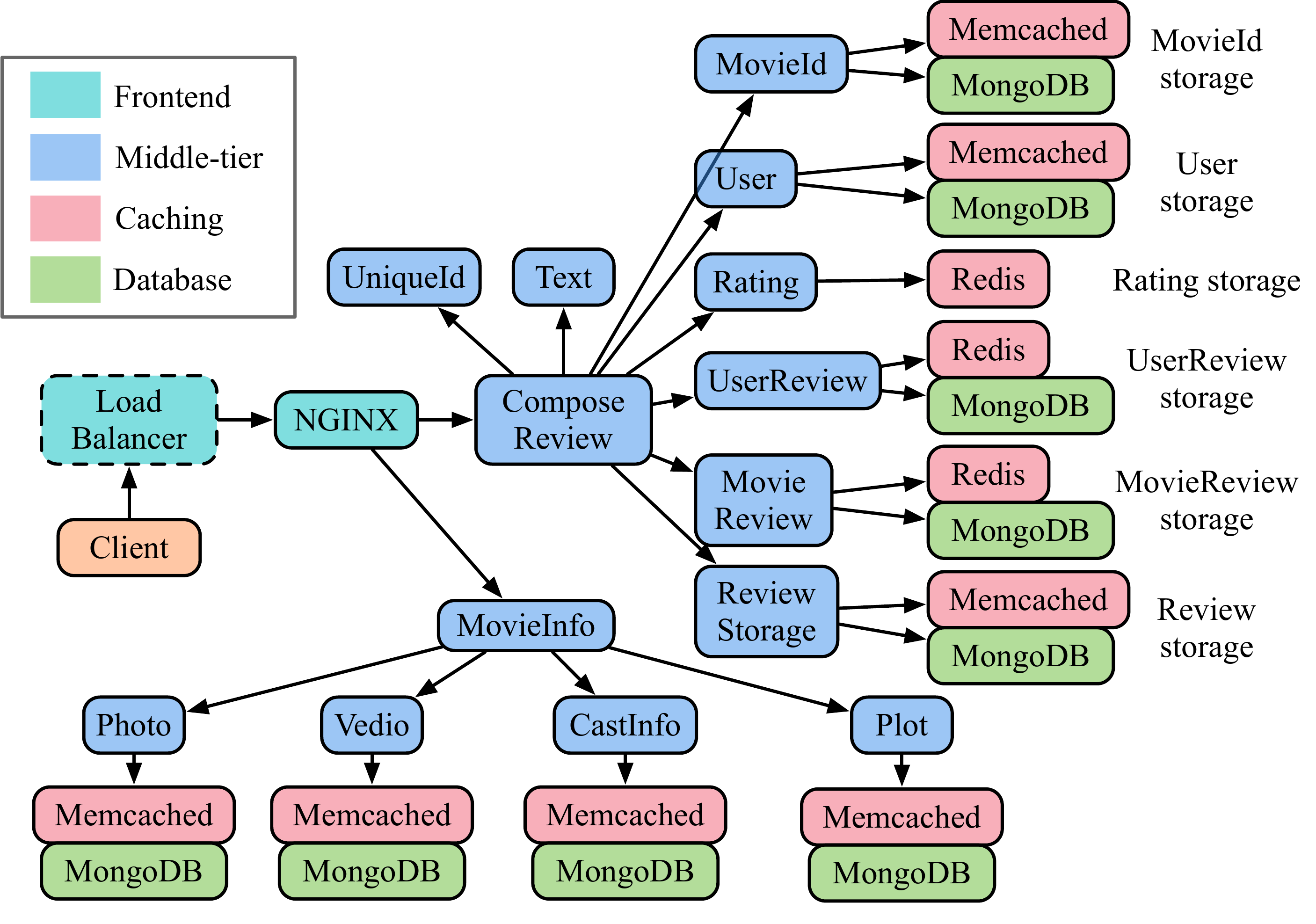}
  \caption{\new{Media Serving microservice architecture~\cite{DeathStarBench} -- \normalfont{client requests first reach a front-end load balancer, which evenly distributes them across the $N$ webserver instances. Then, depending on the type of user request, mid-tiers will be invoked to browse information about a movie, create a new movie review, or get recommendations on movies a user may enjoy. At the right-most of the figure, the requests reach the back-end databases, implemented both with in-memory caching tiers (memcached and Redis), and persistent databases (MongoDB). }}}
  \label{fig:media_service}
\end{figure}

\noindent{\bf{Integrated special-purpose networking hardware:}}
To address the aforementioned issues, a recent line of research proposed to tightly integrate NICs with the processor and memory. In particular, NetDIMM~\cite{NetDIMM} investigates the potential of physical integration of NICs into DIMM memory hardware. However, NetDIMM does not focus on RPC stacks, and it requires designing custom ASICs inside the DIMM hardware, which is hard to achieve at scale in a short term. Another solution to this end is soNUMA~\cite{soNUMA}, which proposes to scale-out coherent NUMA memory interconnects at the datacenter level, therefore bringing cluster machines closer to each other. NeBuLa~\cite{NeBuLa} similarly discusses the implementation of a hardware-offloaded RPC stack on top of soNUMA. NeBuLa introduces a novel mechanism for delivering RPC payloads directly into a processor's L1 cache, and also proposes an efficient in-LLC buffer management and load balancing method that reduces network queueing, and improves the tail latency of RPC requests. However, as with NetDIMM, both NeBuLa and soNUMA require fabrication of custom hardware that is physically integrated with the processor and memory subsystem, and reorganizing the entire datacenter network architecture, which is challenging to achieve at datacenter scale, especially given the frequent design and deployment changes of microservices. Another closely-related work, Optimus Prime~\cite{OptimusPrime}, presents the design of an RPC data transformation accelerator which reduces the CPU pressure from expensive (de)serialization operations common in datacenter communication systems. 

At a high level, Dagger implements an RPC stack, fully offloaded to hardware. The FPGA-based design makes it amenable to the frequent updates present in microservices, and customizable to their diverse needs. In addition, Dagger leverages commercially-available memory interconnects as the interface between the processor and the \textit{Ethernet} NIC, so, in contrast to previous solutions requiring custom hardware and/or specifically designed datacenter networks, our proposed system can be integrated into existing cloud infrastructures without the need for chip tapeouts. We also demonstrate that third party applications, such as memcached can be easily ported to Dagger, with minimal changes to the their codebase. 

\new{Even though Dagger's implementation is based on an FPGA, it is fundamentally different from other FPGA/SmartNIC-related proposals~\cite{msft, NICA, Floem, TONIC} and industrial NICs. All existing networking devices are either ``Smart'' in the way they process packets but remain PCIe-attached, or are more closely-integrated to the main CPU but not programmable, only implementing simple packet delivery functionality. From this perspective, Dagger is the first attempt to leverage FPGAs which are \textit{closely-coupled} with processors over memory interconnects as \textit{programmable networking devices} to implement the networking stack all the way up to the application layer (RPCs). This new approach opens up many system challenges and opportunities to customize networking fabrics to the frequently-evolving interactive cloud services, in a high-performance and efficient manner. In addition, our proposal comes with out-of-the-box support for multi tenancy and co-location. We show how a single physical FPGA adapter can host multiple independent NICs serving different tenants running on the same host. }

%% file: Network_Characterization.tex
\section{Characterizing Networking in Microservices}
\label{sec:network_footprint}

\subsection{Networking Overheads in Microservices}

\begin{figure}[h]
    \vspace{-5pt}
    \centering
    \includegraphics[width=1\linewidth]{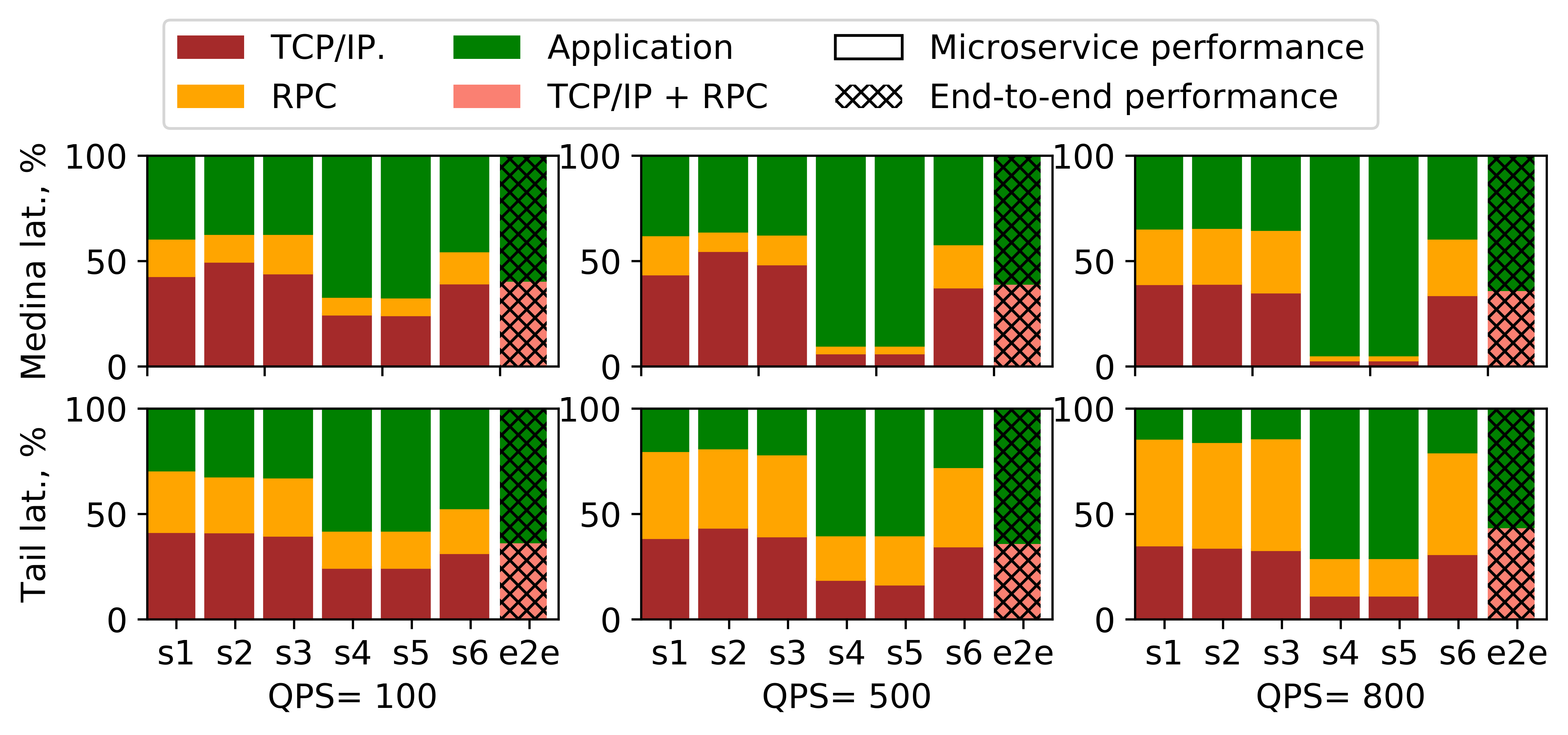}
    \vspace*{-5mm}
  	\caption{Networking as a fraction of median (top) and 99th prc. tail (bottom) latency -- \normalfont{the input load increases progressively from the left-most to the right-most set of figures. The bars denote latency breakdown as recorded in the following individual microservices: s1: Media, s2: User, s3: UniqueID, s4: Text, s5: UserMention, s6: UrlShorten; e2e bar shows the breakdown for the end-to-end latency.} }
    \label{fig:networking_fraction}
\end{figure}

We first study the network characteristics and requirements of interactive, cloud microservices. We use two end-to-end services from the DeathStarBench benchmark suite~\cite{DeathStarBench}, shown in Figures~\ref{fig:social_network} and~\ref{fig:media_service}; a Social Network, and a Media Serving application. We first profile the impact of the RPC stack on the per-tier and end-to-end latency for Social Network, for a representative subset of its microservices, shown in Figure~\ref{fig:networking_fraction}. $s1$ is a \texttt{Media} tier processing embedded images and videos in user posts, $s2$ is the \texttt{User} tier responsible for managing a user's account and interfacing with the backend storage tiers, $s3$ is the \texttt{UniqueID} tier assigning a unique identifier to a new user post, $s4$ represents the \texttt{Text} service to add text to a new post, $s5$ is the \texttt{UserMention} tier to link to another user's account in a post, and finally $s6$ corresponds to the \texttt{UrlShorten} microservice, which shortens links embedded in user posts. 

We break down communication overheads to RPC and TCP/IP processing, and show both median (top) and tail latency (bottom) for different load levels, from the left-most to the right-most figures. Across all tiers, communication accounts for a significant fraction of a microservice's latency, 40\% on average, and up to 80\% for the light in terms of computation \texttt{User} and \texttt{UniqueID} tiers. This also translates to a large fraction of end-to-end application latency going towards networking, despite many microservices overlapping with each other. The latency breakdown for the end-to-end application is shown in the right-most bar of each subfigure in  Figure~\ref{fig:networking_fraction}; communication accounts for at least third of the median and tail end-to-end latency. RPC processing is a large part of communication, and, for some services it is even larger than the TCP/IP stack when looking at tail latency, showing that accelerating the TCP/IP layer alone is not sufficient. A couple of microservices, such as \texttt{Text} and \texttt{UserMention} are more computationally intense, and therefore, their processing time is substantially longer than the communication latency. However, since microservices typically form deep call graphs, long delays in the upstream services can cause request queueing in downstream tiers, incurring cascading QoS violations~\cite{Gan19b}. 

The time devoted to networking increases considerably for higher loads (right plots in Figure~\ref{fig:networking_fraction}), especially when looking at tail latency, due to excessive queueing across the networking stack. Not only does this degrade application performance, but the multiple layers of queueing in the network subsystem introduce performance unpredictability, making it hard for the application to meet its end-to-end QoS target. \new{Note that since we cannot explicitly break down the time between queueing and processing at the application layer, for high loads, e.g., QPS=800, our profiler shows high percentages for RPC processing time; most of this time corresponds to queueing. We observe that the microservice's CPU utilization does not increase proportionally with load, in part because of back pressure from the downstream services. Such aggressive queueing in the RPC layer causes significant memory pressure and high memory interference with other tasks, making tail latency even more unpredictable and high. Overall, Figure~\ref{fig:networking_fraction} shows that networking, both the RPC layer and transport, in microservices is a major performance bottleneck, especially for tiers with a small amount of compute. The results are similar for the other end-to-end services as well}

\subsection{Network Characteristics of Microservices}

\begin{figure}[h]
    \vspace{-5pt}
    \centering
    \includegraphics[width=1\linewidth]{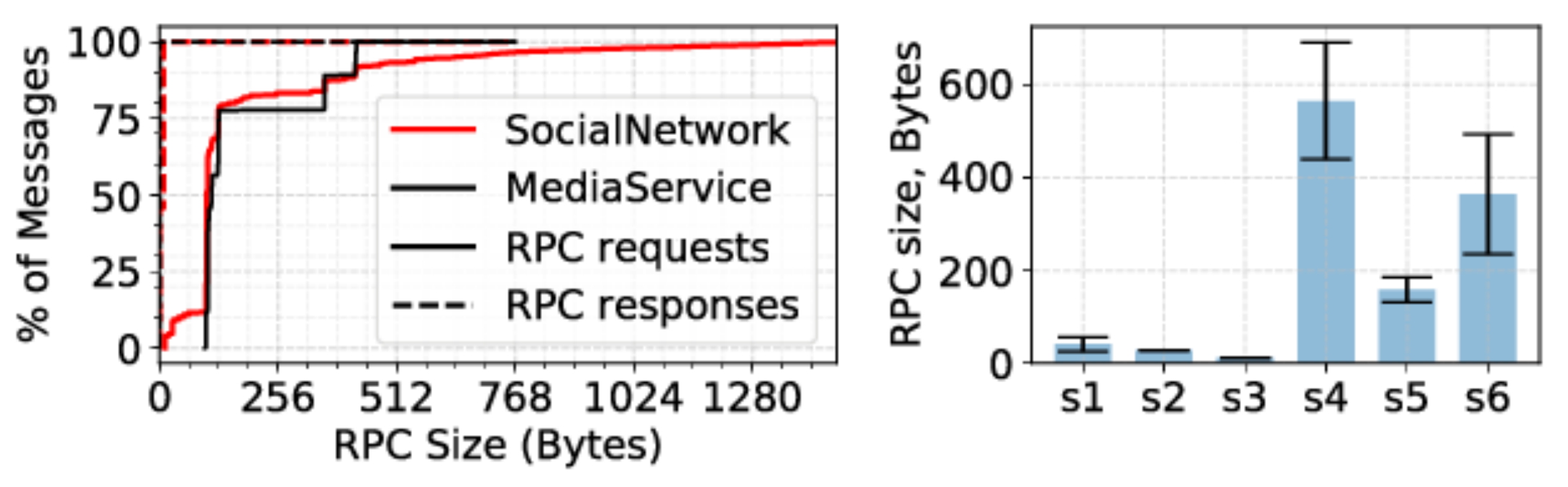}
    \vspace*{-5mm}
  	\caption{Distribution of RPC request and response sizes in Social Network and Media Applications (left); breakdown of RPC sizes for individual microservices (right) -- \normalfont{s1 - s6 are the same as in Figure~\ref{fig:networking_fraction} for Social Network. } }
    \label{fig:rpc_sizes}
\end{figure}

We now analyse the network footprint of microservices. We profile the same Social Network and Media Serving applications~\cite{DeathStarBench}, and show the histogram of their RPC sizes in Figure~\ref{fig:rpc_sizes}. The workload generator follows request distributions representative of real cloud services implemented with microservices~\cite{DeathStarBench}. Specifically, the request mix includes queries for users creating new posts (\textit{Compose Post}), for reading their own timeline (\textit{Read Home Timeline}, or another user's timeline \textit{Read User Timeline}). Depending on the request type, a different subgraph of microservices is invoked~\cite{deathstarbench_github}. 

Figure~\ref{fig:rpc_sizes} (left) shows the cumulative distribution function (CDF) of RPC sizes for each end-to-end service. 75\% of all RPC requests are smaller than 512B. Responses are even more compact, with more than 90\% of packets being smaller then 64B. Commodity networking systems experience poor performance when it comes to transferring small packets due to high per-packet overheads~\cite{HOMA, DBLP:conf/nsdi/KaliaKA19}. This highlights the need for rethinking datacenter networking with the unique properties of microservices in mind. 

Additionally, Figure~\ref{fig:rpc_sizes} (right) shows that different microservices in the same application have very different RPC sizes. 
For example, the median RPC size in the \textit{Text} service is 580B, while the \textit{Media}, \textit{User}, and \textit{UniqueID} services never have RPCs larger than 64B. Such wide variation in RPC sizes across microservices shows that ``one-size-fits-all'' is a poor fit for microservice networking. Balancing between optimizing for large versus small packets is a long standing problem in networking. To achieve the best for this trade-off, Dagger employs reconfigurable hardware acceleration using FPGAs, to ensure that the networking stack can be tailored to the needs and characteristics of a given set of active microservices.

\subsection{CPU Contention between Logic and Networking}

\begin{figure}[h]
    \vspace{-5pt}
    \centering
    \includegraphics[width=1\linewidth]{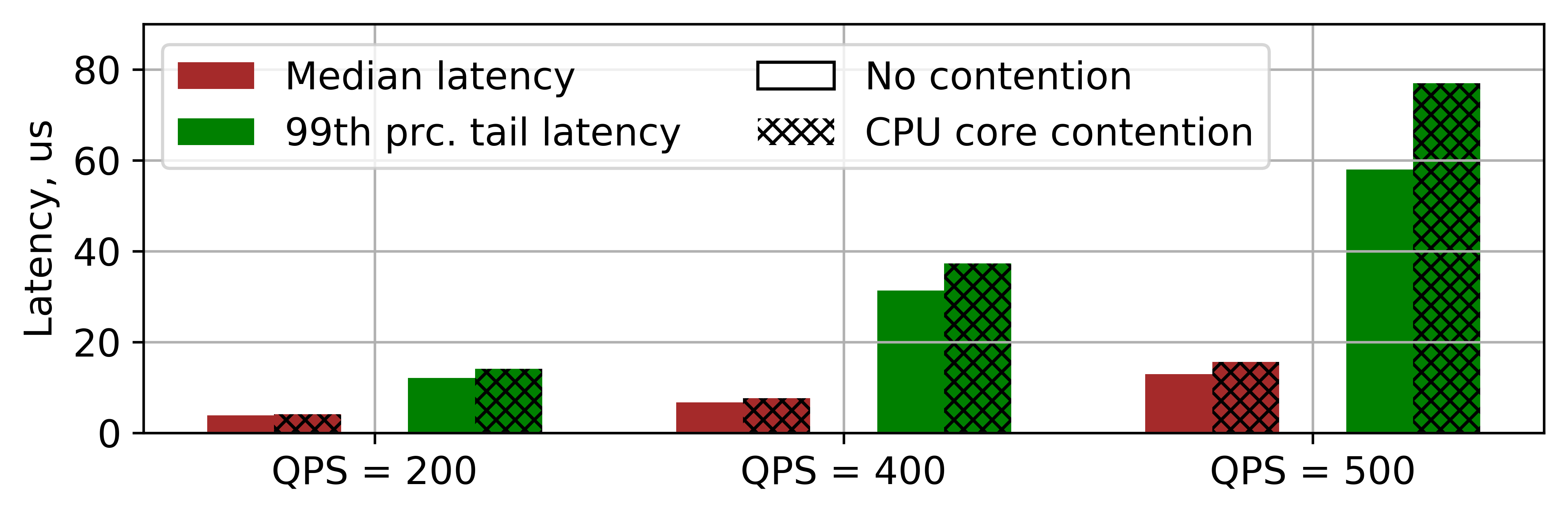}
    \vspace*{-5mm}
  	\caption{Impact of the CPU resource interference between networking and application logic on end-to-end tail latency -- \normalfont{solid bars show latency when network processing and application logic run on different physical cores, while shaded bars show latency when network processing and application logic threads share the same CPU cores. } }
    \label{fig:resource_contention}
\end{figure}

The small size of microservices means that a server can concurrently host a large number of microservices, promoting resource efficiency. On the negative side, multi-tenancy also introduces resource interference, especially when CPU cores are used for both microservice logic and network processing. To quantify the resource contention between application logic and networking, we bind network interrupt service routines for each of the NIC's queues to a fixed set of N logical CPU cores ($N = 4$ in this experiment). We then run the Social Network service: (1) on the other $N$ CPU cores within the same NUMA node so that their execution does not interfere with networking, and (2) on the same $N$ cores as networking to observe the impact of interference. The resulting end-to-end request latency in each case is shown in Figure~\ref{fig:resource_contention}.

The experiment shows that when both application logic and network processing contend for the same CPU resources, end-to-end latency (both median and tail) suffers. As expected, interference becomes worse as the system load increases, especially when it comes to tail latency, which is more sensitive to performance unpredictability. Note that in addition to the network interrupt handling layer, RPC processing also interferes with application processing. Since the RPC stack is technically a part of the application logic in the service's default implementation, we are not able to isolate its impact; however, the high resource contention of Figure~\ref{fig:resource_contention} already justifies the need for offloading the networking stack from the host processor. Dedicating, alternatively, cores specifically for network processing is not resource efficient, since network load fluctuates over time, dedicated networking cores are still prone to CPU-related overheads, and they can introduce interference in the last level cache (LLC) and main memory subsystems.

This analysis highlights three unique requirements for networking systems aimed at microservice deployments. First, RPC processing should be offloaded to dedicated hardware, to avoid CPU-related overheads and interference with the application logic. Second, systems should be optimized for small requests/responses, which dominate the network traffic of microservices. Finally, communication frameworks should be programmable to handle the diverse needs of microservices, and adjust to their frequent changes.

%% file: Dagger_Design.tex
\section{Dagger Design}
\label{sec:dagger}

\subsection{High-Level Architecture and Design Principles}

\begin{figure}[h]
\centering
\vspace{-5pt}
\begin{subfigure}{1\linewidth}
    \includegraphics[width=1\linewidth]{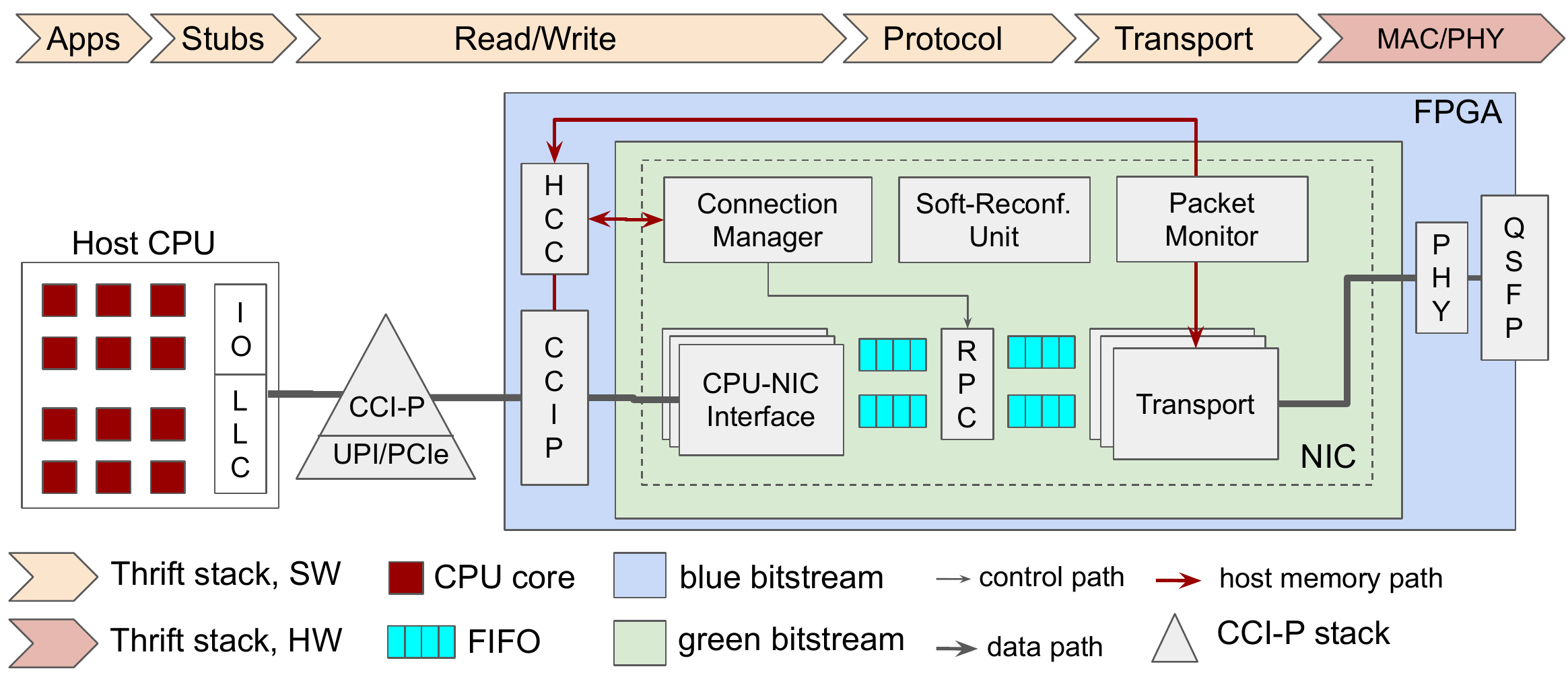}
    \label{fig:top_level_diagram}
\end{subfigure}%

\begin{subfigure}{1\linewidth}
    \includegraphics[width=1\linewidth]{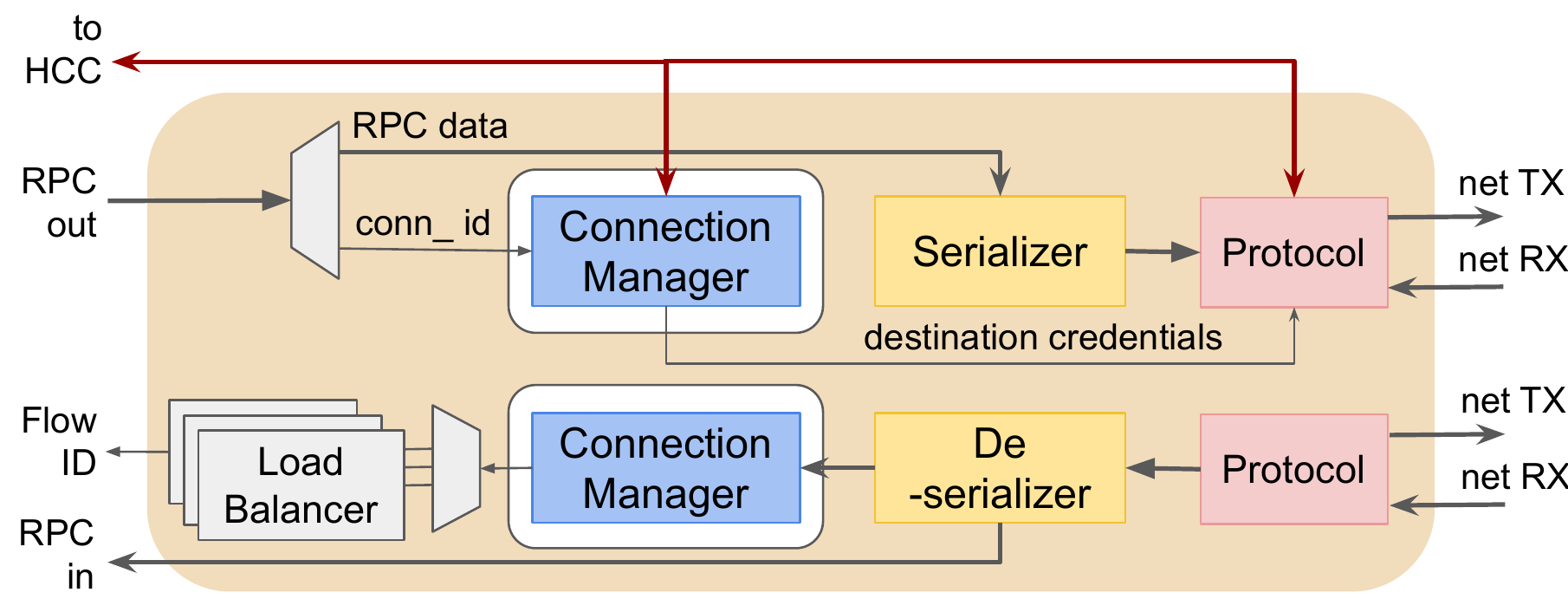}
    \label{fig:rpc_unit}
\end{subfigure}%
 \vspace{-12pt} 
\caption{\new{Top-level architecture of Dagger (top) and  zoom-in of the PRC unit (bottom) -- \normalfont{the bar on top shows approximate mapping of the Dagger stack onto the Thrift RPC~\cite{Thrift} stack. }}}
\label{fig:top_level}
\end{figure}

We design Dagger with the unique network properties and requirements of microservices in mind, discussed in Section~\ref{sec:network_footprint}. Although Dagger is optimized for microservices, it is still beneficial for traditional interactive cloud applications, as we will show in Section~\ref{sec:evaluation}. Dagger's top-level architecture is shown in Figure~\ref{fig:top_level}. Dagger is based on three main design principles: i) \textit{full hardware offload}, ii) \textit{tight coupling}, and iii) \textit{reconfigurability}. 

First, Dagger offloads the entire RPC stack in hardware, to eliminate all interference between application logic and network processing in the host CPU. The remaining software components of Dagger running on the host CPU are lightweight, and are only responsible for the connection set-up and for exposing the RPC API to applications. The software stack implements the API with zero-copying and directly places incoming RPC requests and responses to dedicated buffers (queues, rings) accessible by the hardware which is synthesized on an FPGA (shown by the NIC module in Figure~\ref{fig:top_level}). The rest of the processing is handled by the NIC.

The second design principle of Dagger is tightly coupling the hardware-accelerated network fabric with the host CPU. In contrast to existing high-performance and/or programmable NICs, which leverage PCIe-based interfaces, Dagger uses NUMA memory interconnects as the interface with the host processor to optimize the transfer of small fine-grained RPCs, by piggybacking on the hardware coherence protocol. The NUMA memory interconnect (Intel UPI in our current implementation) is encapsulated into the CCI-P protocol stack, as shown by the triangle in Figure~\ref{fig:top_level}. CCI-P~\cite{CCIP} is a protocol stack between the host CPU and the FPGA designed by Intel, which in addition to UPI, also wraps-up two PCIe links.

Finally, the design of Dagger focuses on reconfigurability by leveraging an FPGA as the physical medium. Our current design is based on the Intel OPAE HDK that defines two regions of programmable logic: the blue and the green bitstreams. The former is fixed in the FGPA configuration and is not exposed to the FPGA users. This includes the implementation of system components, such as the CCI-P interface IP, Ethernet PHY, clock generation infrastructure, etc. The blue bitstream is managed by the cloud providers, and undergoes periodic updates. The green bitstream is used to implement the user logic and is fully programmable, with a pre-synthesized bitstreams. We implement the Dagger NIC in the green region, as shown in Figure~\ref{fig:top_level}. Our design is modular and configurable: different hardware parameters and components can be selected via SystemVerilog macros/parameters and synthesized. We call this \textit{hard configuration}, and use it only for coarse-grained control decisions. For example, Dagger supports multiple different CPU-NIC interfaces; the choice of the specific scheme is performed via hard configuration, by selecting the corresponding IP blocks and configuring the design with them. Similarly, the choice of the transport layer, sizes of on-chip caches (e.g. connection caches), flow FIFOs, etc., are also enabled by hard configuration. Since hard configuration requires preparing dedicated bitstreams, and it incurs overheads to reprogram the FPGA, some fine-grained control decisions are still made via soft configuration. Soft configuration is based on soft register files accessible by the host CPU via PCIe MMIOs, and the corresponding control logic (Soft-Reconfiguration Unit in Figure~\ref{fig:top_level}). Dagger uses soft configuration to control the batch size of CCI-P data transfers, provision the transmit and receive rings, configure their number and sizes, configure the number of active RPC flows, choose a load balancing scheme, etc. Soft reconfiguration comes with certain logic overheads, however, it enables fast and fine-grained tuning of various parameters of the Dagger framework at runtime. 

The remaining blocks shown in Figure~\ref{fig:top_level} (top) are the Connection Manager used for setting up connections and storing all connection-related metadata, the Packet Monitor that collects various networking statistics, and auxiliary components, such as FIFOs, for proper synchronization of different blocks in the RPC pipeline. The Host Coherent Cache (HCC) is another important auxiliary unit in Dagger. HCC is a small (128 KB) direct-mapped cache implemented in the blue bitstream, which is fully coherent with the host's memory, via the CCI-P stack. HCC is used to hold cache connection states and the necessary structures for the transport layer on the NIC, while the actual data resides in the host memory. This way we avoid requiring dedicated DRAM hardware for the FPGA. This makes NIC cache misses cheaper compared to PCIe-based NICs, since the CCI-P stack provides hardware support for data consistency between the host DRAM and the HCC.

\subsection{\new{Dagger API and Threading Model}} \label{sec:threading_model}

\new{The API is designed following the standard client-server architecture of cloud applications, and is inspired by the Thrift RPC~\cite{Thrift} framework as well as the Google Protocol Buffers interface, primarily for compatibility reasons, since many microservices rely on either of these APIs. Other RPC APIs, such as gRPC~\cite{gRPC} or Finagle~\cite{finagle} can also be supported by the design. Similarly to commercial RPC stacks, Dagger comes with its own Interface Definition Language (IDL) and code generator. We adopt the Google Protobuf IDL for Dagger; an example of our interface definition scheme is shown in Listing~\ref{d_idl_listing}.} \vspace{0.1in}

\definecolor{keywordcolor}{HTML}{cc33ff}

\lstdefinestyle{customcpp}{
aboveskip=0in,
belowskip=0in,
abovecaptionskip=0.08in,
belowcaptionskip=0.2in,
captionpos=b,
xleftmargin=\parindent,
language=C++,
morekeywords={GetRequest, GetResponse, KeyValueStore, SetRequest, SetResponse, int32},
showstringspaces=false,
basicstyle={\small\linespread{0.6}\fontseries{sb}\normalsize\ttfamily},
keywordstyle=\bfseries\color{keywordcolor},
commentstyle=\itshape\color{green!40!black},
}

\begin{lstlisting}[style=customcpp, label=d_idl_listing, caption=Dagger IDL on an example of a KVS service.]
Message GetRequest {  Message GetResponse {
  int32 timestamp;      int32 timestamp;
  char[32] key;         char[32] value;
}                     }

Service KeyValueStore {
  rpc get(GetRequest) returns(GetResponse);
  rpc set(SetRequest) returns(SetResponse);
}
\end{lstlisting}

\new{The code generator parses target IDL files and produces client and server stubs which wrap up the low-level RPC structures being written/read to/from the hardware into the high-level service API function calls. The latter defines two main classes: the \textit{RpcThreadedServer} and the \textit{RpcClientPool} for each client-server pair. The \textit{RpcClientPool} encapsulates a pool of RPC clients (\textit{RpcClient}) that concurrently call remote procedures registered in the corresponding \textit{RpcThreadedServer} as \textit{RpcServerThread} objects wrapping server event loops and dispatch threads. Dagger supports both asynchronous (non-blocking) and synchronous (blocking) calls. In the former case, each \textit{RpcClient} contains the associated \textit{CompletionQueue} object which accumulates completed requests. The \textit{CompletionQueue} might also invoke arbitrary continuation callback functions upon receiving RPC responses, if so desired.}

\begin{figure}[h]
    \centering
    \setlength{\belowcaptionskip}{-10pt}
    \includegraphics[width=1\linewidth]{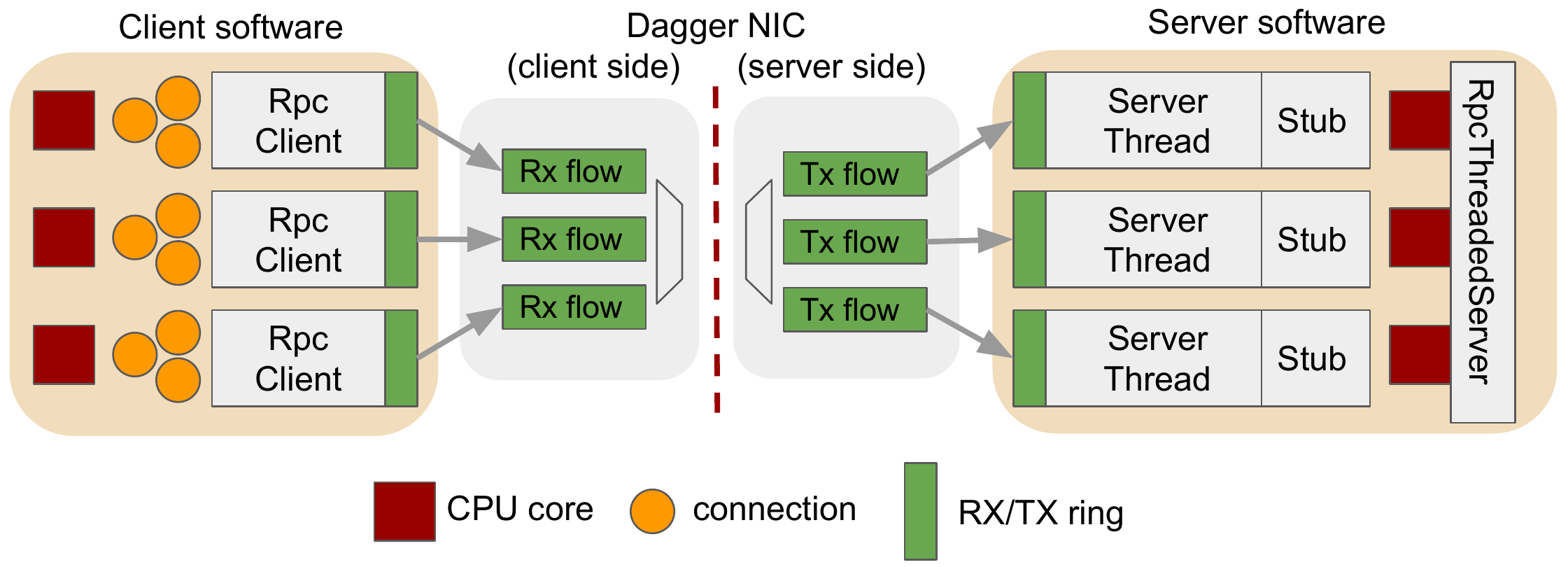}
    \vspace*{-15pt}
  	\caption{\new{Dagger threading model -- \normalfont {the green Rx/Tx flows correspond to hardware flows on the NIC (only single direction flows are shown for brevity; in the real system, each NIC runs both Rx and Tx flows).}}}
    \label{fig:threading_model}
\end{figure}

\new{The Dagger threading model is co-designed across hardware and software and is fully configurable, as is shown in Figure~\ref{fig:threading_model}.}

\new{Dagger provisions multiple flows (queues) on the NIC, such that each flow is 1-to-1 mapped to the corresponding RX/TX ring in software. The rings themselves are 1-to-1 mapped to \textit{RpcClient}'s and \textit{RpcServerThread}'s. The number of NIC flows, and therefore RX/TX rings, determines the degree of concurrency in the Dagger hardware, and is programmable via hard configuration. Note that the number of flows need not necessarily be equal to the number of CPU cores. However, in the basic scheme, shown in Figure~\ref{fig:threading_model}, the number of NIC flows is decided based on the number of logical CPU cores, such that each core gets a dedicated parallel flow on the NIC. Similarly to FaRM, Dagger runs RPC handlers in dispatch threads to avoid inter-thread communication overheads. Also, with a small change in software, it can be configured to run RPC handlers in separate worker threads if required for long-running RPCs; this does not require any hardware changes. }

\new{Our threading model allows opening an arbitrary number of connections on each \textit{RpcClient}. In this case, the connections on a certain \textit{RpcClient} share the same RX/TX ring, so following the RDMA terminology, Dagger implements the Shared Receive Queue (SRQ) model~\cite{MPI}. Note that with the programmable threading model, Dagger can be configured to run in a single flow mode with a single RX/TX ring shared between multiple CPU cores. This enables models similar to~\cite{RPCValet} which target addressing load imbalance. At the other extreme, provisioning flows and rings on a per-connection basis is also possible, although such a scheme scales poorly and suffers from high load imbalance. }

\new{Dagger manages connections entirely in hardware which further reduces CPU load and improves the look-up of connection information for active flows. The NIC includes the Connection Manager (CM) module, as shown in Figure~\ref{fig:top_level}. The connection table interface maps connection IDs (c\_id) onto tuples <src\_flow, dest\_addr, load\_balancer>. The src\_flow field specifies the ID of the flow receiving requests from the client. The NIC reads this information to ensure that the responses are steered to the same flows where requests came from. The dest\_addr and load\_balancer fields define the address of the destination host and preferred load balancing scheme for requests withing this connection. }

\new{The CM is designed as a simple direct-mapped cache with specific memory organization. In order to make the cache access concurrent and avoid stalls in the RPC flows, the cache breaks the above interface tuple into three tables indexed by the $\ceil[\big]{log(N)}$ LSBs (where N is the table size) of the connection ID providing 1W3R functionality. This is required because at the same time (cycle), three independent hardware agents might read from the cache: the RPC outgoing flow (to get the destination credentials), the incoming flow (to get the flow or load balancer), and the CM itself (to open and close connections). The size N of the cache is adjustable with hard configuration and can be chosen based on the expected number of connections the application might open. If some application requires many connections, N can be set to a high value giving more connection cache space to this application in favor of other NIC memory structures. Given the available size of FPGA on-chip memory (53Mb total minus 8.8Mb in the green region) and the size of the current connection tuple (8-12B)x3, the FPGA can be configured to cache at most 153K connections; sufficient for most application scenarios. In addition, the connection cache can be easily backed by DRAM (either externally attached to the FPGA or by the host DRAM) to allow more connections with certain performance penalty due to NIC cache misses. Although this functionality is not yet implemented in our current design, we plan to integrate it as part of future work (see the red lines in Figure~\ref{fig:top_level}).}

\subsection{NUMA Interconnects as NIC Interfaces} \label{sec:memory_interconnects}

PCIe links have acted as the default NIC I/O interfaces for the past several decades. Despite the bus being a standard peripheral interconnect in any modern processor, a lot of prior work has shown that PCIe is not efficient as a NIC I/O interface~\cite{Cambridge, stanford_mmio}. The inefficiency is mainly introduced in the transmission path, when the NIC is fetching network packets from the host memory. In the simplest case, commercial NICs use DMA transfers initiated by MMIO doorbell transactions to read packet descriptors and payloads from the software buffers; an approach known as the \textit{doorbell method}~\cite{rdma_1}. 

However, the na\"ive doorbell scheme experiences inefficiencies when targeting small requests. The MMIO transactions are slow, mainly because they are implemented as non-cacheable writes, and expensive: every MMIO request should be explicitly issued by the processor. To reduce the overhead of MMIOs, modern high-performance NICs, such as Mellanox RDMA NICs, implement doorbell batching~\cite{rdma_1}, an optimizations that allows grouping multiple requests into a single DMA transaction initiated with a single MMIO. While this solution noticeably increases the performance of doorbells, it still relies on MMIO messages and is only applicable when requests can be aggressively batched, which is not always possible for latency-sensitive flows. Another proposal~\cite{stanford_mmio} suggests eliminating DMAs, and transferring data only using MMIOs when requests fit in the MMIO's MTU, usually 1 cache line. This improves latency since data are transferred within a single transaction, however, performance is still limited by the low throughput of MMIOs, and during high load, this can overload the processor. 

The fundamental limitation of PCIe protocols is that their design is primarily geared towards Producer-Consumer dataflow models~\cite{CXL}. The standard doorbell model works well under streaming flows and large data transfers. However, RPC requests do not always conform to such patterns. As we showed in  Section~\ref{sec:network_footprint}, RPC sizes in microservices, and in other datacenter applications~\cite{HOMA}, are small, ranging from a few bytes and up to few kilo-bytes. In addition, the strict latency requirements of interactive services often disqualify batching, forcing NICs to handle fine-grained data chunks rather than streaming flows. This issue is further exacerbated when RPC frameworks do not just send requests, but also involve some amount of data processing. For example, Thrift RPC was designed to work with complex data objects that are not uniform in memory; for example, they might contain nested structures and references to other objects. In this case, RPCs must be (de)serialized~\cite{OptimusPrime}, with existing PCIe models being very inefficient in fetching such non-uniformly placed objects. The standard doorbells used in all PCIe-attached NICs require expensive and CPU-inefficient data transformations before sending data to the NIC~\cite{OptimusPrime}. 

The main insight in leveraging memory interconnects as the NIC I/O is that they allow data transfers to be handled entirely in hardware. The memory consistency state machines (NUMA cache coherence protocols) implemented as a part of the processor's memory subsystem are designed to provide efficient and fast data flows between coherent agents; processors, or more generally, NUMA nodes. Making the NIC act as another NUMA node would allow it to closely integrate its I/O into the processor's memory subsystem, therefore providing a pure hardware CPU-NIC interface without the need for explicit notifications of data updates from the processor. This improves the CPU efficiency of sending small RPCs, since the only operation the processor needs to do is write the RPC requests/responses to the buffer it shares with the NIC, with the actual transfer handled entirely by the interconnect state machines. This increased CPU's efficiency significantly improves per-core RPC throughput (Section~\ref{sec:evaluation}).

The exact scheme of data movements inside coherent busses depends on the specific NUMA interconnect model. Some specifications, such as the upcoming peripheral memory interconnect CXL~\cite{CXL}, allow non-cacheable writes to the device memory, meaning that the CPU can directly write RPCs to the NIC, so in addition to improved CPU efficiency, the model also reduces latency, since only one bus transaction is required to send data to the device.

Note that we do not compare the \textit{physical} performance of PCIe with respect to memory busses in this work. The peak bandwidth of both interconnects is implementation specific and depends on the number of lanes in the physical layer and the generation of the interconnect. Moreover, some memory interconnects are based on PCIe, so they use the same physical medium with the same bandwidth. All sources of performance gain shown in this work come from the difference in the \textit{logical} upper-level communication models enabled by PCIe vs NUMA protocols. However, the theoretical bandwidth of the UPI bus which we use as the memory interconnect goes up to 19.2 GB/s; slightly higher than the 15.74 GB/s of the PCIe Gen3x16.

\subsection{Implementation of the NIC Interface} \label{sec:cpu_nic_interface}

\begin{figure}[h]
    \vspace{-5pt}
    \centering
    \setlength{\belowcaptionskip}{-10pt}
    \includegraphics[width=1\linewidth]{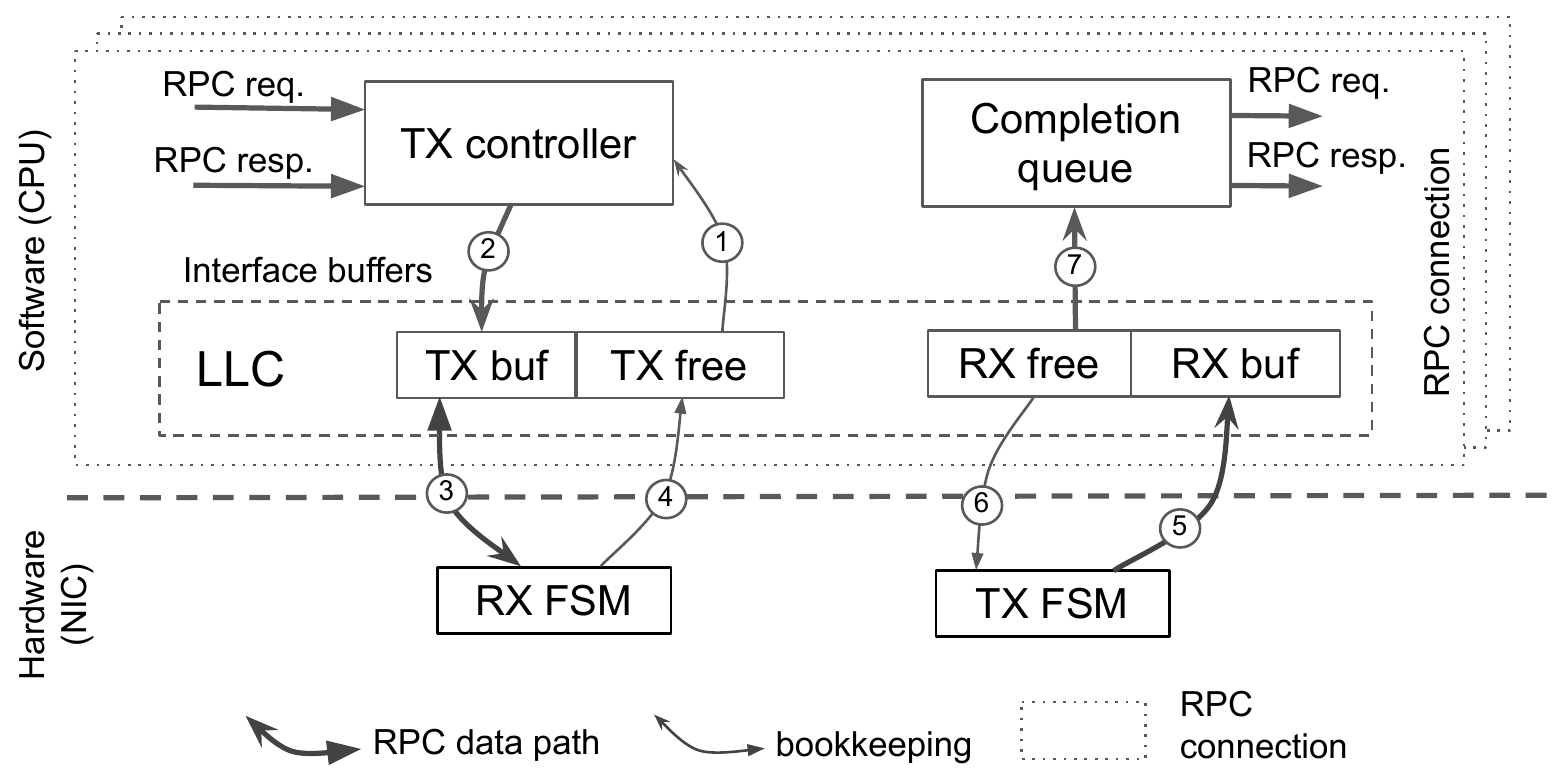}
  	\caption{CPU-NIC interface diagram -- \normalfont {RX path: TX controller writes new RPC requests/responses {\large \textcircled{\small 2}} to a free entry in the TX buffer {\large \textcircled{\small 1}}. The RX FSM on the FPGA fetches RPC objects from the TX buffer over CCI-P {\large \textcircled{\small 3}}; it also does the bookkeeping {\large \textcircled{\small 4}} to release previously-fetched entries. TX path: the TX FSM puts newly-received RPC objects to the RX buffer, and asynchronously fetches the next free entries during bookkeeping {\large \textcircled{\small 6}}. The RPC payload is finally copied to the completion queue {\large \textcircled{\small 7}}.}}
    \label{fig:cpu_nic_interface}
\end{figure}

The NIC I/O interface consists of the receiving and transmitting paths as seen from the NIC. The CPU-NIC interface diagram is shown in Figure~\ref{fig:cpu_nic_interface}.

Dagger provisions network buffers (RX/TX rings) on per-NIC flow basis. Each RX/TX pair reflects the communication channel between a single \textit{RpcClient} and the corresponding \textit{RpcThreadedServer} wrapping the dispatch thread. Such buffer provisioning enables lock-free access to the rings from the client and server threads~\cite{NeBuLa}. As stated above, the same \textit{RpcClient} can run multiple threads (each within a separate connection) therefore making them to share the corresponding RX/TX rings. In that case, explicit locking in the \textit{RpcClient} RX/TX path is required to ensure consistent data transfer.

The RX/TX ring are used for incoming and outgoing RPCs, respectively. The stack is symmetric, the same architecture serves both requests and responses, i.e., the same NIC and the software stack can be used for both RPC clients and servers. Request types are distinguished by the \textit{request type} field that is a part of every RPC packet. RX/TX rings are comprised of RX/TX buffers and free buffers. The former store RPC payloads for all requests until the NIC/completion queue acknowledges receiving the data by placing the ID of the corresponding RX/TX buffer entry into the free buffer. The size of the TX rings is determined by the rate of incoming RPCs and the time it takes for the NIC to fetch data and is configured during the NIC initialization. In our current prototype, the CCI-P-based memory interconnect, based on Intel UPI, delivers data from the software buffers to the NIC within 400 ns with another 400 ns required for sending back the bookkeeping information. The CCI-P bus can support up to 128 outstanding requests before reaching its bandwidth limit, so Dagger sends multiple asynchronous requests, while the bookkeeping information of in flight requests is pending. The size in the number of requests of the TX rings per flow is $\ceil[\big]{Thr_{per\_flow} * 0.8 / 10^6}$, where $Thr_{per\_flow}$ is the desired throughput of the flow. For $Thr_{per\_flow}$ = 12.4 Mrps (the maximum throughput a single CPU core can sustain in our system), the size of the TX rings should be at least $10\times$ the mean RPC size to avoid flow blocking. Given the typical sizes of RPCs in microservices, every flow requires 0.64-12.8 KB of TX buffers. 

The RX rings accumulate a batch of requests before sending them to the completion queue. Therefore, the RX buffer size is equal to $B\times$ the mean RPC size, where $B$ is the width of CCI-P batching. In our experiments, the maximum sustained throughput is achieved with batching of $B=4$. Below we detail the hardware design of the receiving and transmitting paths in Dagger. 

\subsubsection{Receiving path (RX)}

We implement three standard PCIe-based methods for fetching RPC data from the processor: doorbell, doorbell batching, and MMIO-based transfer, also known as WQE-by-MMIO in~\cite{rdma_1}, alongside with our proposed method based on memory interconnects. The doorbell transfer is based on standard PCIe DMAs initiated via MMIO transactions. We use the DMA engine provided by the Intel OPAE HDK that operates over CCI-P. The doorbell batching is implemented by grouping CCI-P DMAs into single transactions and initiating the entire batch with a single doorbell. In the MMIO-based method, RPC requests and responses are transferred by writing them to a shared buffer allocated as an MMIO memory region on the FPGA. A typical MMIO write on most processors and NICs is a Write-Combine transaction. Write-Combine is used to avoid generating multiple PCIe transactions when writing data of a cache line size (64B), since processors normally commit word-aligned store instructions (8B). A Write-Combine buffer accumulates multiple stores and sends cache line-long chunks when the buffer is full or is being explicitly flushed. We modify this scheme with parallel store instructions based on the AVX-256 ISA extension and we do not use Write-Combining; this allows us to further reduce the MMIO latency. In this mode, Dagger writes every 64B of RPC requests using two \_mm256\_store\_si256 stores. 

The memory interconnect-based interface in Dagger is implemented over the Intel UPI coherent bus. Unfortunately, since CCI-P is the first commercially-available implementation of the UPI bus on an FPGA, the corresponding IP core, which is a part of the blue region of the FPGA bitstream and is therefore not accessible to users, only allows accessing CPU memory via memory polling. Given this limitation, Dagger starts by polling its local cache which is coherent with the processor's LLC and relies on invalidation messages to bring new data from software buffers. However, since the FPGA allocates data in its local cache in this case, it causes the CPU to lose ownership of the corresponding cache lines therefore hurting the data transfer's efficiency. For this reason, Dagger dynamically switches to direct polling of the processor's LLC when the load becomes high, as defined by a programmable threshold. Note that the current limitation of transferring data through polling is not fundamental to Dagger's design, but rather an implementation artifact of the currently-available realization of a memory interconnect on an FPGA. The next generation of FPGAs (Intel Agilex) already integrates a more advanced interconnect IP based on the CXL~\cite{CXL} specification. As mentioned above, the CXL protocol supports direct data writes from CPUs to the FPGA's memory without the need for memory polling. 

\subsubsection{Transmitting path (TX)} \label{sec:tx_path}

\begin{figure}[h]
    \vspace{-5pt}
    \centering
    \setlength{\belowcaptionskip}{-10pt}
    \includegraphics[width=1\linewidth]{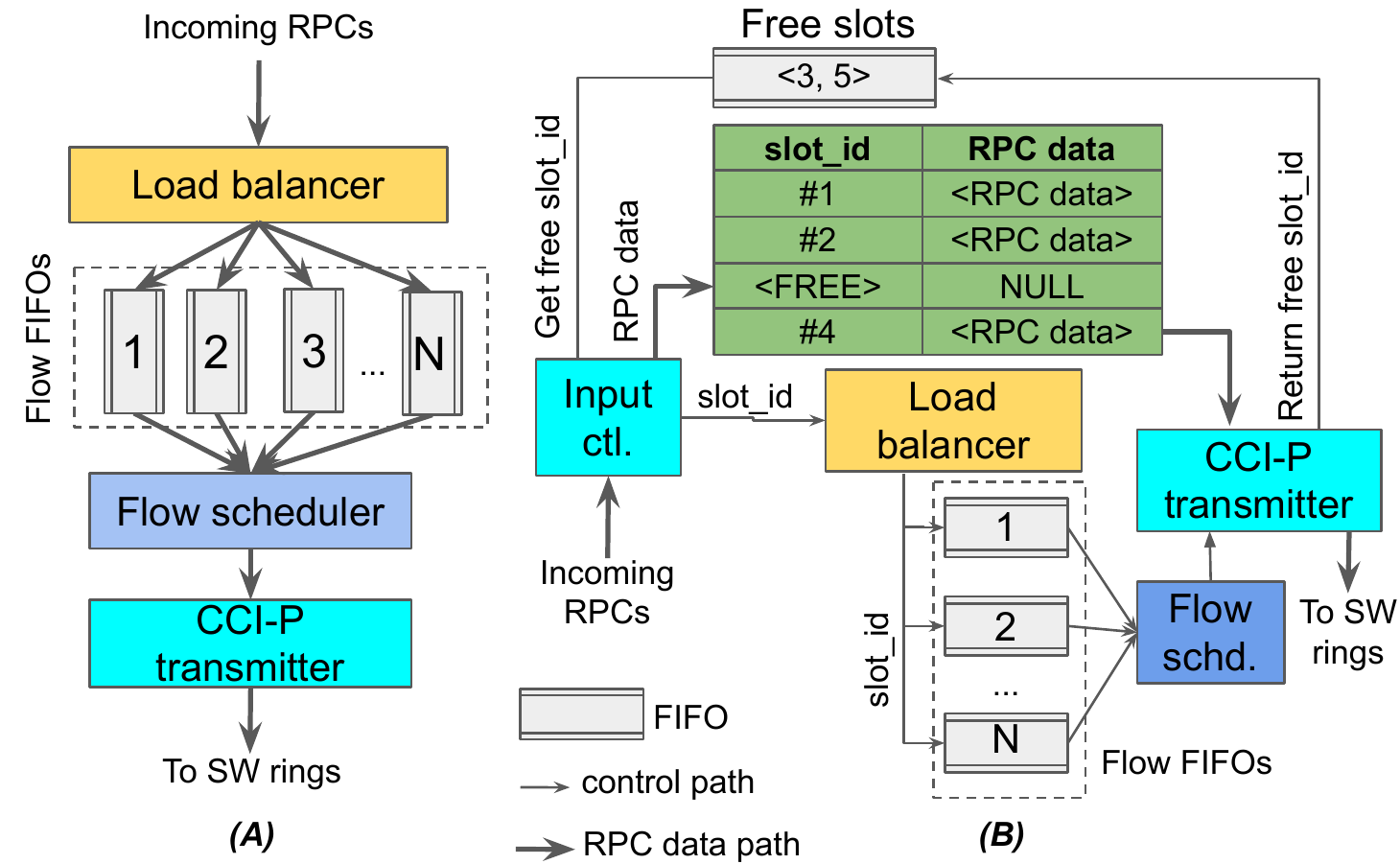}
  	\caption{(A) Architecture and (B) Implementation of TX path. }
    \label{fig:tx_paths}
\end{figure}

The transmission path has some additional complexity compared to the RX path due to the need for load balancing and scheduling/distribution of flows over the active queues. Figure~\ref{fig:tx_paths}A shows the architecture of Dagger's TX path. 

Incoming RPCs should be distributed across the available receiving rings (RX buffers in Figure~\ref{fig:cpu_nic_interface}), and within a ring, the requests can be batched for more efficient data transfer, if a service's latency target permits it. Distribution of requests across rings is handled by the Load Balancer (as a part of the RPC unit, refer to Figure~\ref{fig:top_level} (bottom)) that directs incoming RPCs to the corresponding Flow FIFOs. Each RX buffer gets a dedicated Flow FIFO on the NIC. The Load Balancer currently supports two request distribution schemes: dynamic uniform steering and static load balancing. In the first case, incoming RPCs are evenly distributed across the available flows. In the static balancing, the RPCs are distributed based on the information stored in the corresponding connection tuple on the server. In addition, we leave some room in the design for implementation of application-specific load balancers (e.g. the Object-Level core affinity mechanism in MICA~\cite{MICA}). The Flow Scheduler then picks a Flow FIFO that already contains enough requests to form a transmission batch and instructs the CCI-P transmitter to send the batch to the corresponding ring. 

The architecture in Figure~\ref{fig:tx_paths}A is implemented in hardware, as shown in Figure~\ref{fig:tx_paths}B. Since Dagger's RPCs are at least 64 Bytes long, storing them in FIFOs for each flow and then multiplexing the flows is not practical, and adds complexity to the design. Instead, Dagger implements a request buffer, shown as the green table in Figure~\ref{fig:tx_paths}, which stores all incoming RPCs in a lookup table indexed by the slot\_id. The Free Slot FIFO is designed to keep track of free entries in the request buffer. The Flow FIFOs in this case only contain references (slot\_ids) to the actual RPC data in the table. When sending data, the CCI-P transmitter directly reads RPC payloads from the request table based on the references read from the corresponding Flow FIFO. The size of the request table is equal to $B * N_{flows}$ entries, where B is the CCI-P batching and $N_{flows}$ is the total number of NIC flows. 

\subsection{RPC Pipeline} \label{sec:rpc_pipelne}

The CPU-NIC interface is the first unit of the RPC pipeline; the other two being the RPC module and Transport layer (Figure~\ref{fig:top_level}). The architecture of the RPC module is zoomed-in in Figure~\ref{fig:top_level} (bottom). It implements request serialization/de-serialization between ready-to-use RPC objects, as read from the processor and the network. If compression/encryption is required, the corresponding logic can optionally be integrated into the unit. In addition, the RPC module also contains a set of load balancers that decide which flow to steer incoming requests. The choice of the load balancer is controlled at server granularity, i.e. each server can specify the load balancer it requires when registering connections on it. The Protocol is the last module of the RPC unit. It is designed to implement RPC-optimized protocol layers such as congestion control, piggybacking acknowledgement, transactions built into the RC stack, etc., and is currently idle. Systems similar to TONIC~\cite{TONIC} can be used to implement the Protocol unit.

The RPC unit is connected (over FIFOs, for synchronization) with the Transport layer which implements a version of the UDP/IP protocol and sends outgoing serialized RPC requests to the Ethernet network. Since data transformations and RPC transport protocols are not the focus of this work, we simplify these parts of the pipeline. Our current implementation only supports RPCs with continuous arguments that do not contain references to other objects and application-specific data-structures requiring custom serialization, and the Protocol unit is currently idle - it simply forwards all packets to the network. In the following-up work, we plan to extend Dagger with reliable transports and with RPC-specific congestion control mechanisms which has been shown to be more efficient in datacenter networks than TCP~\cite{DBLP:conf/nsdi/KaliaKA19}.

\subsection{Dagger Implementation} \label{sec:implementation}

We implement Dagger on an Intel Broadwell CPU/FPGA hybrid architecture. The host processor is a server-class Intel Xeon E5-2600v4 CPU integrated with an Arria 10 GX1150 FPGA. The hardware part (Dagger NIC) is written in SystemVerilog using the Intel OPAE HDK library. The software part is designed in C++11 and is compiled by GCC under under the O3 optimization level. The Dagger IDL code generator is written in Python 3.7. The software modules of our RPC stack run in user space, and the NIC buffers are allocated by the FPGA driver in the application virtual address space. The most important implementation parameters of the Dagger NIC design are summarized in Table~\ref{tab:implementation_details}.

\begin{table}[]
\begin{threeparttable}
  \caption{Implementation specifications of Dagger NIC.}
  \label{tab:implementation_details}
\begin{tabular}{l||l}
\hline
\multicolumn{1}{c||}{\textbf{Parameter}}       & \textbf{Value} \\ \hline \hline
CPU-NIC interface clock frequency, MHz         & 200 - 300            \\ 
RPC unit clock frequency, MHz                  & 200            \\ 
Transport clock frequency, MHz                 & 200            \\ 
Max number of NIC flows \tnote{1} & 512  \\ 
FPGA resource usage, LUT (K) \tnote{2}            &   87.1 (20\%)             \\ 
FPGA resource usage, BRAM blocks (M20K)\tnote{2}           &   555 (20\%)             \\ 
FPGA resource usage, registers (K) \tnote{2}             &      120.8          \\ \hline
\end{tabular}
\begin{tablenotes}\footnotesize
\item[1] Assuming 65K entries in the connection cache and ensuring the FPGA BRAM and logic utilization do not exceed 50\%
\item[2] Including the blue region; UPI-based NIC I/O with 64 NIC flows and 65K entries in the connection cache
\vspace{-8pt} 
\end{tablenotes}
\end{threeparttable}
\end{table}

\subsection{Limitations} \label{sec:limitations}

An important limitation of the current design  is the lack of support for efficient RPC reassembling to enable transfer of requests larger than the cache line size. In contrast to PCIe DMA, memory interconnects implement relaxed memory consistency models, which is one of the key reasons behind their efficiency. Therefore, the MTU of a typical memory interconnect is only a single cache line~\cite{NeBuLa}. A na\"ive solution to address the issue of sending larger RPCs is to reassemble requests in software. However, this will introduce CPU overheads and violate our first design principle. Another solution, as proposed in NeBuLa, leverages Content Addressable Memory (CAM) for on-chip reassembling in hardware. Unfortunately, CAMs are expensive in terms of area and energy, and it is challenging to implement them with low overheads on an FPGA. Efficient RPC reassembling in hardware is a challenging issue, and we plan to address it as part of future work. As of now, Dagger only features software-based RPC reassembling.

%% file: Evaluation.tex
\section{Evaluation}
\label{sec:evaluation}

\subsection{Methodology}

We evaluate Dagger along five dimensions. First, we compare Dagger with prior work on efficient RPC processing based on user-space networking and RDMA. Second, we evaluate different CPU-NIC interfaces and show the performance benefits of memory interconnects over PCIe. Third, we show how Dagger scales with the number of CPU threads. In addition, we demonstrate that Dagger can be easily integrated with existing datacenter applications, such as memcached and MICA, offering dramatic latency improvement under realistic workloads. Finally, we run a simple microservice application on top of Dagger showing that our RPC stack is indeed suitable for multi-tier systems. Table~\ref{tab:platform_spec} shows the specification of the hardware platform used.

\begin{table}[h]
  \caption{Hardware specifications of experimental platform. }
  \label{tab:platform_spec}
\begin{tabular}{llll}
\hline
\multicolumn{4}{l}{\textbf{CPU: Intel Xeon E5-2600v4}}                                                                        \\ \hline
 & Cores               &  & \begin{tabular}[c]{@{}l@{}}12 cores (OOO), 2 threads per core, \\ 2.4 GHz, 14 nm\end{tabular}               \\
 & LLC                 &  & 30720 kB, 64 B                                                                                    \\
 & Additional features &  & AVX-2,  DDIO, VT-x                                                                                \\
 & OS                  &  & \begin{tabular}[c]{@{}l@{}}CentOS Linux,\\ Kernel Linux 3.10.0\end{tabular}                       \\ \hline
\multicolumn{4}{l}{\textbf{Interconnect: CCI-P: 2x PCIe and 1x UPI}}                                                          \\ \hline
 & PCIe                &  & \begin{tabular}[c]{@{}l@{}}Gen3x8, 7.87 GB/s, 2 links,\\ total bandwidth 15.74 GB/s\end{tabular}  \\
 & UPI                 &  & \begin{tabular}[c]{@{}l@{}}9.6 GT/s (19.2 GB/s), 1 link,\\ total bandwidth 19.2 GB/s\end{tabular} \\ \hline
\multicolumn{4}{l}{\textbf{FPGA: Arria 10  GX1150}}                                                                           \\ \hline
 & Max frequency       &  & 400 MHz                                                                                           \\ \hline
\end{tabular}
\end{table}

Due to the limitation on the number of FPGA-enabled machines on the Intel vLab cluster, we instantiate two identical Dagger NICs on the same FPGA and connect them to each other via a loop-back network. We then give the NICs fair round-robin access to the CCI-P bus by multiplexing it. Note that since the main contribution of Dagger is in CPU-NIC interface and the RPC pipeline, the absence of physical networking does not affect our findings.

\subsection{Performance Comparison across RPC Platforms}

We compare the performance of Dagger's RPC acceleration fabric with four related proposals, based on DPDK user-space networking IX~\cite{IX}, raw user-space networking eRPC~\cite{DBLP:conf/nsdi/KaliaKA19}, RDMA FaSST~\cite{FASST}, and the in-memory integrated NIC NetDIMM~\cite{NetDIMM}. Table~\ref{tab:eval_3} shows the median round trip latency and the throughput achieved by each system. We also show the TOR (Top of Rack) delay assumed in each work, the size of the being transferred objects, and their type. Note that if the object type is ``msg'', this means that the system does not implement the RPC layers of the networking stack, and the reported results do not include the overhead of RPC processing.

\setlength\belowcaptionskip{-1ex}
 \begin{table}[h!]
 \begin{threeparttable}
\centering
  \caption{Median round trip time (RTT) and throughput of single-core RPCs compared to related work \tnote{1}. }
  \label{tab:eval_3}
\begin{tabular}{c|c|c|c|c||c}
\hline
 &
  \textbf{IX} &
  \textbf{\begin{tabular}[c]{@{}c@{}}FaSST\end{tabular}} &
  \textbf{\begin{tabular}[c]{@{}c@{}}eRPC\end{tabular}} &
  \textbf{\begin{tabular}[c]{@{}c@{}}Net-\\DIMM\end{tabular}} &
  \textbf{\begin{tabular}[c]{@{}c@{}}Dagger\end{tabular}} \\ \hline \hline
\textbf{Objects} &
  \begin{tabular}[c]{@{}c@{}}64B\\ msg\end{tabular} &
  \begin{tabular}[c]{@{}c@{}}48B\\ RPC\end{tabular} &
  \begin{tabular}[c]{@{}c@{}}32B\\ RPC\end{tabular} &
  \begin{tabular}[c]{@{}c@{}}64B\\ msg\end{tabular} &
  \begin{tabular}[c]{@{}c@{}}64B\\ RPC\end{tabular} \\ 
\textbf{\begin{tabular}[c]{@{}c@{}}TOR \\ delay\end{tabular}} & N/A  & 0.3 us & 0.3 us & 0.1 us & 0.3 us \\ 
\textbf{\begin{tabular}[c]{@{}c@{}}RTT, us\end{tabular}}         & 11.4 & 2.8 & 2.3 & 2.2 & 2.1  \\ 
\textbf{\begin{tabular}[c]{@{}c@{}}Thr., \\ Mrps\end{tabular}}   & 1.5 & 4.8 \tnote{2} & 4.96 \tnote{2} & N/A & 12.4  \\ \hline
\end{tabular}
\begin{tablenotes}\footnotesize
\item[1] Performance numbers are provided from corresponding papers
\item[2] Recorded in symmetric experimental settings
\end{tablenotes}
\end{threeparttable}
\end{table}

As seen from Table~\ref{tab:eval_3}, Dagger shows $1.3 - 2.5\times$ (depending on experimental settings) higher per-core RPC throughput than the RDMA-based solution, FaSST, and the DPDK-based eRPC. The gain partially comes from offloading the entire RPC stack on hardware and leaving only a single memory write in the critical RPC path on the processor. In addition, approximately 14\% of performance improvement is enabled by replacing the doorbell batching model with our memory interconnect-based interface. Note that neither PCIe's nor UPI's physical bandwidth is saturated in this experiment, so these 14\% come from the better messaging model enabled by memory interconnects. Table~\ref{tab:eval_3} also shows that Dagger achieves the lowest median round trip time of $2.1us$, while significantly improving throughput compared to both user-space and kernel-level networking. This is better than FaSST, and even the integrated solution NetDIMM, and is comparable with eRPC.

\subsection{Comparison of CPU-NIC Interfaces}

\begin{figure}[h]
    \vspace{-5pt}
    \centering
    \includegraphics[scale=0.43]{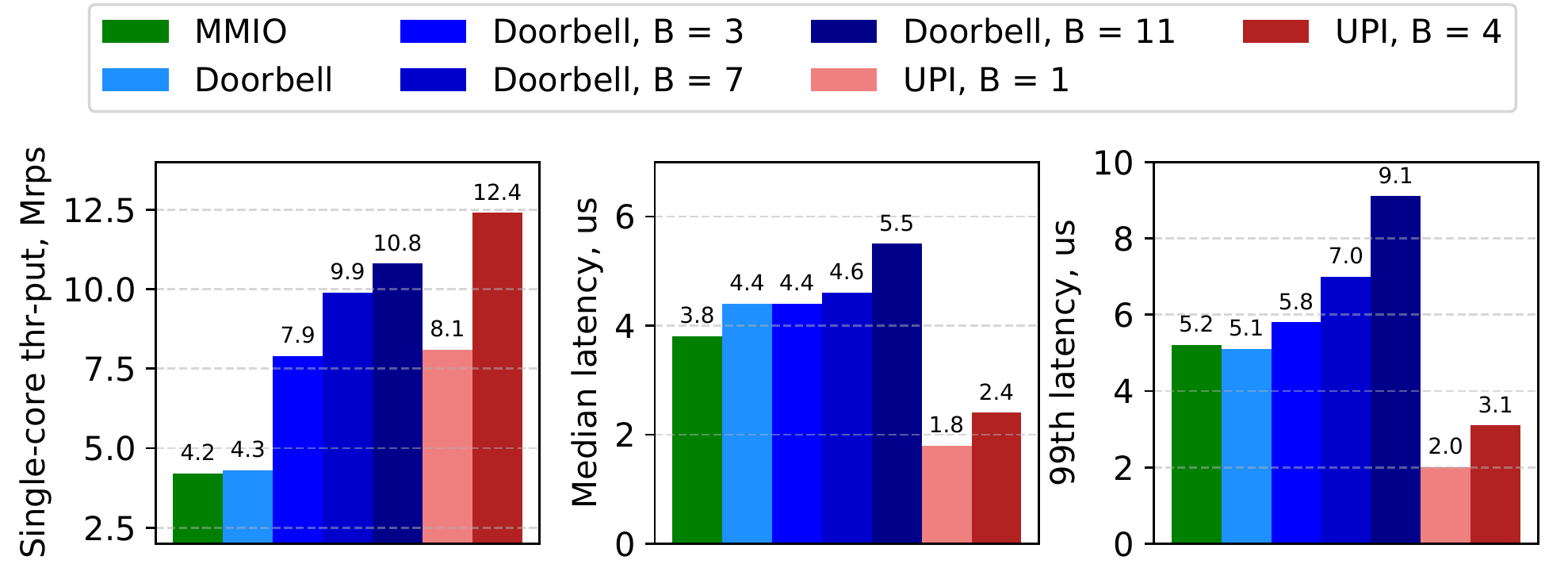}
    \vspace*{-5mm}
  	\caption{Dagger's single-core throughput and latency for different CPU-NIC interfaces (RX path) when transferring 64 Bytes RPCs -- \normalfont {B denotes request batching. }}
    \label{fig:eval_1}
\end{figure}

Figure~\ref{fig:eval_1} shows the comparison of Dagger's end-to-end single-core latency and throughput for different CPU-NIC interfaces. Unless otherwise specified, the PCIe CPU-NIC interface is based on a single PCIe Gen3x8 link. The maximum theoretical (physically bounded) throughput is 122 Mrps for 64 Byte RPCs on all cores.

As seen from Figure~\ref{fig:eval_1}, the lowest median and tail latency over a PCIe bus is achieved when the RX path uses MMIO writes, which is reasonable, since in this case all RPCs are being written within a single PCIe transaction. However, the method fails to deliver high throughput, with the best reported result being 4.2 Mrps. A similar throughput of 4.3 Mrps but with higher median latency is reported when Dagger is using non-batched doorbells showing that their performance is limited by the rate of initiating MMIOs. The only way to increase the efficiency of doorbells is to use batching which enables reaching a throughput of 10.8 Mrps for batch of $B = 11$. The memory interconnect-based transfer (UPI in Figure~\ref{fig:eval_1}) achieves single-core throughput of 12.4 Mrps with $B = 4$, and demonstrates noticeably lower median and tail latency. Note that the latency improvement does not exclusively come from the reduced number of bus transactions for the transfer using the memory interconnect. We conduct another experiment in which we access an address in the shared memory over PCIe (using DMA) and over UPI. The PCIe DMA gives us $450 us$ of median one-way latency while the UPI read achieves $400 us$. This shows that UPI is physically slightly faster than PCIe. \new{Finally, we measured the maximum single-core throughput of Dagger at 16.5 Mrps, with best-effort request processing by allowing arbitrary packet drops by the server. }

\subsection{Latency vs Throughput}

\begin{figure}[h!]
  \vspace{-8pt}
  \centering
  \begin{minipage}[b]{0.49\linewidth}
    \centering
    \includegraphics[width=\linewidth]{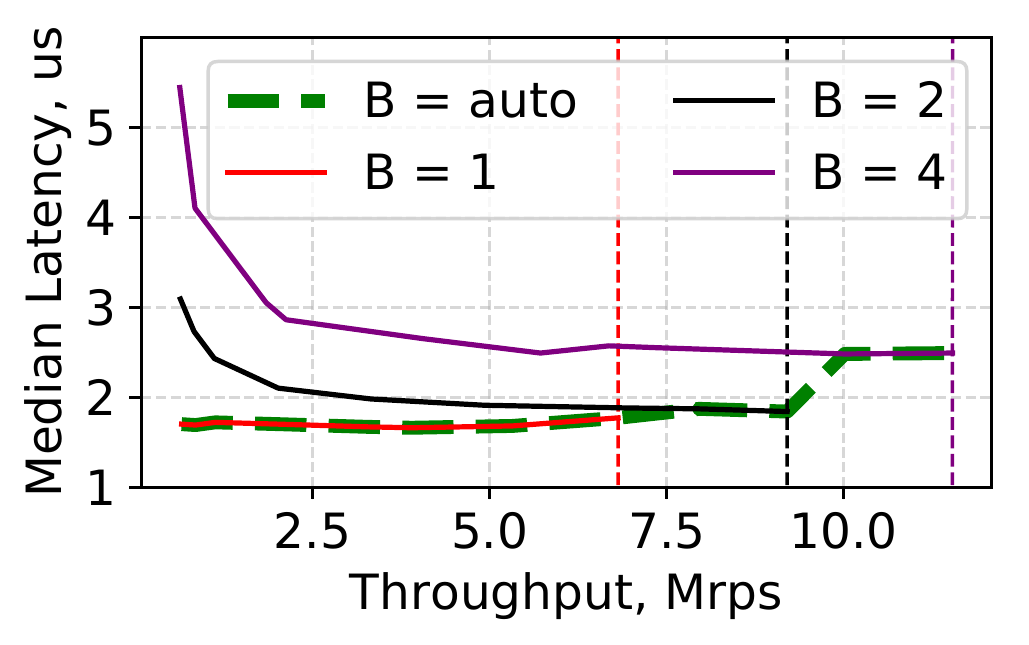}
  \end{minipage}
  \hfill
  \begin{minipage}[b]{0.49\linewidth}
    \centering
    \includegraphics[width=\linewidth]{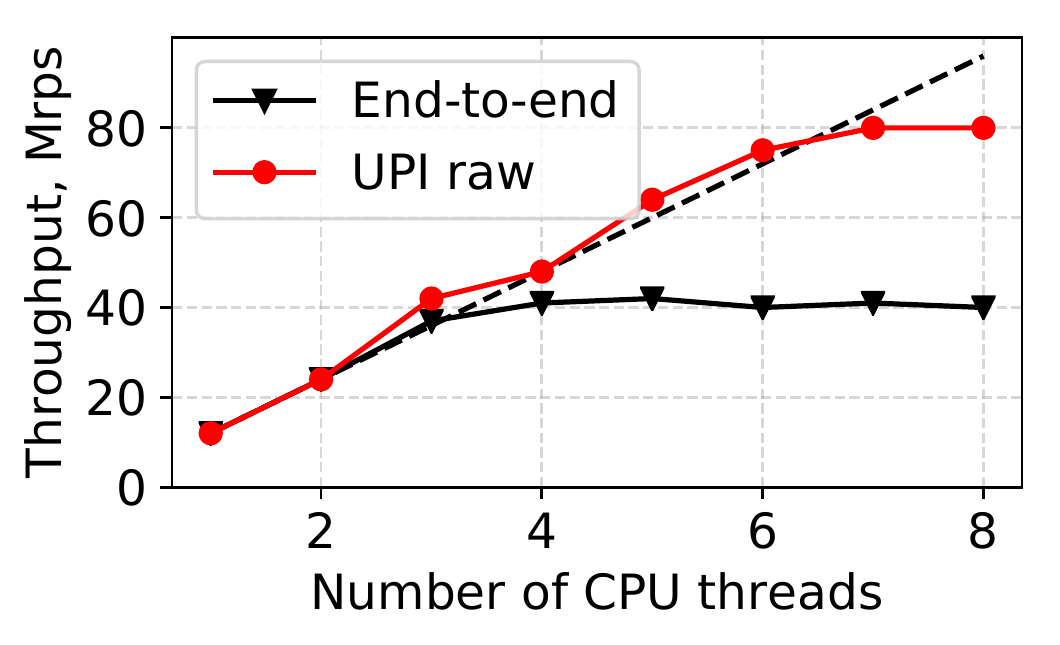}
  \end{minipage}
    \caption{Latency-Throughput curves for single-core asynchronous round-trip 64B RPCs (left) and multi-core scalability of sending 64B requests (right) -- \normalfont {B denotes batching, dotted lines show the saturation point. The black line shows end-to-end RPC throughput, the dashed line denotes estimated linear scalability, and the red line shows the results of the UPI bus's scalability with raw idle requests. }}
    \label{fig:eval_2}
\end{figure}

Figure~\ref{fig:eval_2} (left) shows Dagger's latency under different loads with the memory interconnect-based NIC. 

\setlength\belowcaptionskip{-1ex}

With CCI-P batching $B = 1$, Dagger achieves the lowest median latency (round trip time) of $1.8 us$, which remains stable across the entire load range, up to the throughput saturation point of 7.2 Mrps. When increasing batching to $B = 4$, Dagger's throughput increases to 12.4 Mrps with a latency of $2.8 us$. Note that when the load is low, the request latency is relatively high since the RPC pipeline needs to wait until the batch is full. Dagger leverages soft configuration to adjust the batch size dynamically when the load becomes high so that the throughput advantages of batching do not come at a latency cost (as shown by the green dashed line in Figure~\ref{fig:eval_2} (left)).

\subsection{Thread Scalability}

Figure~\ref{fig:eval_2} (right) plots the scalability of Dagger with the number of CPU threads (logical cores). The system throughput scales linearly up to 4 threads (2 physical cores) and remains flat at 42 Mrps. Note that since we run both the RPC client and server on the same CPU, this effectively translates to 84 Mrps as seen by the processor. This result is $\approx23\%$ higher than the performance reported in the FaSST paper on the CIB cluster, and is $3.5\times$ better than IX under the same number of cores. The saturation results signal that the current bottleneck is not the processor. The NIC itself, which is capable of processing up to 200 Mrps, is also far from saturated. To better understand the scalability of Dagger, we run another experiment in which we send idle memory read requests over the UPI interconnect, the results of which are shown in Figure~\ref{fig:eval_2} (right) in red. The throughput of idle memory reads also scales linearly up to 80 Mrps with 7 threads, and stays flat when one more thread is added. Note that the UPI bus has the total physical bandwidth of 19.2 GB/s (Table~\ref{tab:platform_spec}) which is significantly more than 80Mrps for 64 Byte RPCs. Based on this experiment, we conclude that the current bottleneck is the implementation of the UPI end-point on the FPGA in the blue region. Since this region is encrypted, we are not able to optimize it in the current prototype, however, the upcoming generation of Intel Agilex FPGAs will have a dedicated hard IP core for memory interconnects, which should address this scalability issue. 

\subsection{End-to-End Evaluation on KVS systems} \label{sec:kvs_eval}

We evaluate the end-to-end performance of Dagger on two real KVS systems: memcached~\cite{memcached} and MICA~\cite{MICA}. Memcached is a popular in-memory key-value store, widely used in microservices~\cite{DeathStarBench,Sriraman2018SA,OptimusPrime}. In this experiment, we run the original version of memcached over Dagger with the UPI-based I/O instead of the native transport protocol based on TCP/IP for SET and GET commands. We modify only $\approx$50 LOC of the Memcached source code in order to integrate it with Dagger. We also keep the original memcached protocol to verify the integrity and correctness of the data.

\new{While memcached is a suitable application because of its wide adoption in both industry and academia, it is relatively slow ($\approx$12$\times$ slower than Dagger). Therefore, the performance of memcached over Dagger is bottlenecked by memcached itself, which does not allow Dagger to achieve its full potential. In order to make the evaluation more comprehensive, we also port MICA over Dagger, another well-known KVS system in the academic community. MICA is designed specifically to provide high throughput for small requests, and, in contrast to memcached, it can be, under certain workloads, network-bounded. We run MICA over Dagger with no changes to the original codebase; we simply implement a MICA server application which integrates it with Dagger with $\approx$200 LOC.}

\begin{figure}[h!]
  \vspace{-6pt}
  \centering
  \begin{minipage}[b]{0.49\linewidth}
    \centering
    \includegraphics[width=\linewidth]{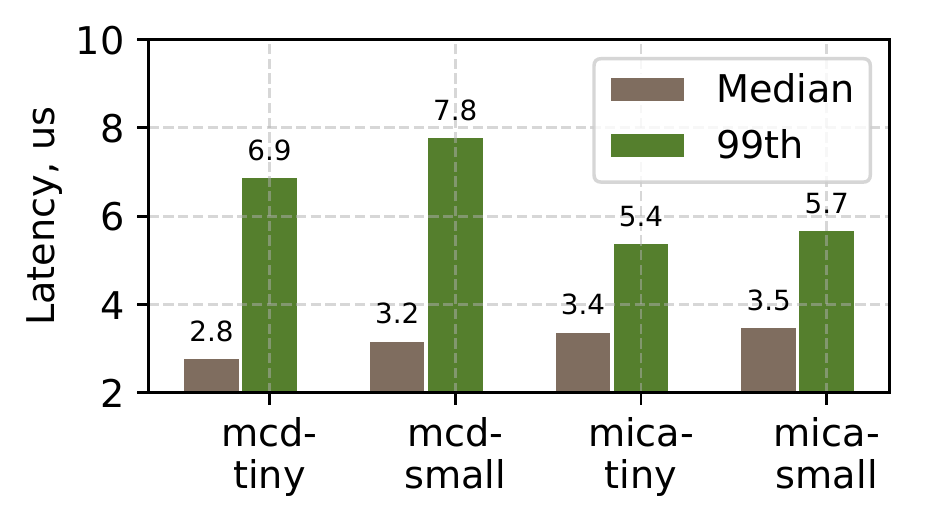}
  \end{minipage}
  \hfill
  \begin{minipage}[b]{0.49\linewidth}
    \centering
    \includegraphics[width=\linewidth]{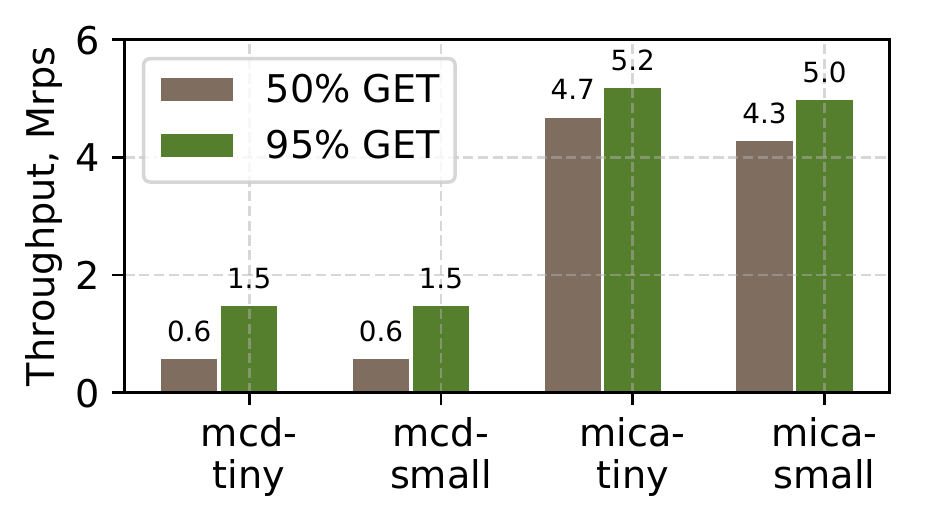}
  \end{minipage}
  \vspace{-6pt}
    \caption{\new{Performance of memcached and MICA running over Dagger: request latency (left) and throughput (right) -- \normalfont {latency results recorded under the write-intensive workload and the peak single-core throughput. }}}
    \label{fig:eval_4}
\end{figure}

\new{To evaluate the KVS systems, we generate two types of datasets similar to the ones used to evaluate MICA~\cite{MICA}: tiny (8B keys and 8B values) and small (16B keys and 32B values). We populate both memcached and MICA KVS with 10M and 200M unique key-value pairs respectively, and access them over the Dagger fabric, following a Zipfian distribution~\cite{zipf} with skewness of 0.99. For MICA, we use their original benchmark and workload generator~\cite{MICASource}. We load both systems with two types of workloads: write-intense (set/get = 50\%/50\%) and read-intense (set/get = 5\%/95\%). We adjust the workload generator such that the number of packet drops on the server is always < 1\%. The results of running memcached and MICA with Dagger on a single core are shown in Figure~\ref{fig:eval_4}.}

\setlength\belowcaptionskip{-1ex}

\new{Figure~\ref{fig:eval_4} (left) shows the latency of the KVS systems running over Dagger. Latency is defined as the round-trip time between the moment when a request is issued by a core and when the result is received. We measure latency for the write-intensive (set/get = 50\%/50\%) workload in a similar way to~\cite{MICA} to ensure a fair comparison. Memcached achieves the lowest median latency of $2.8 us$, and a $99^{th}$ percentile latency of $6.9 us$, when running under a $0.6 Mrps$ load on a single core for tiny requests. For small requests, the results are slightly higher. The MICA KVS shows a better tail latency of $5.4 us$ and $5.7 us$ for tiny and small requests, respectively. It is $4.4 - 5.2\times$ lower than the latency numbers reported in the original MICA work when running it over the lightweight DPDK-based networking stack. This result shows that Dagger can offer dramatic latency reduction in high performance KVS systems compared to optimized user-space networking. }

\new{The KVS throughput results are shown in Figure~\ref{fig:eval_4} (right). With the workload that we use, the systems are still bottlenecked by the key-value store. Dagger can reach up to $12.4 Mrps$ of single-core throughput, where memcached and MICA had a limit of $0.6 - 1.6$ and $4.8 - 7.8$ Mrps, respectively. For this reason, the single-core throughput improvement of Dagger remains hidden. At the same time, these results show that integrating our RPC stack does not add any additional throughput overhead compared to software-based networking systems. To better load MICA, we also test it under a distribution with skewness of 0.9999, which yields even higher data locality, and therefore better cache utilization. With such a workload, Dagger achieves a throughput of 10.2 Mrps and 9.8 Mrps for read- and write-intensive workloads with the same latency numbers, as in Figure~\ref{fig:eval_4}, therefore bringing the performance of MICA closer to the peak performance of Dagger. }

\new{We do not show results of multi-core scalability for MICA, since the extensive amount of LLC contention introduces considerable instability in the results. The contention is caused by running both client and server on the same CPU, without the possibility to partition the cache, therefore allowing them to share the same portion of LLC. The client runs the workload generator which reads 1.49GB of data at a very high rate from its internal buffer making the LLC traffic very high. As part of future work, we plan to deploy Dagger to a cluster environment with physically distributed FPGAs to avoid client-server colocation, and measure more representative multi-core throughput results. }

\vspace{-5pt}

\subsection{End-to-End Evaluation on Microservices} \label{sec:eval_microservices}

\begin{figure}[h]
    \centering
    \includegraphics[width=1\linewidth]{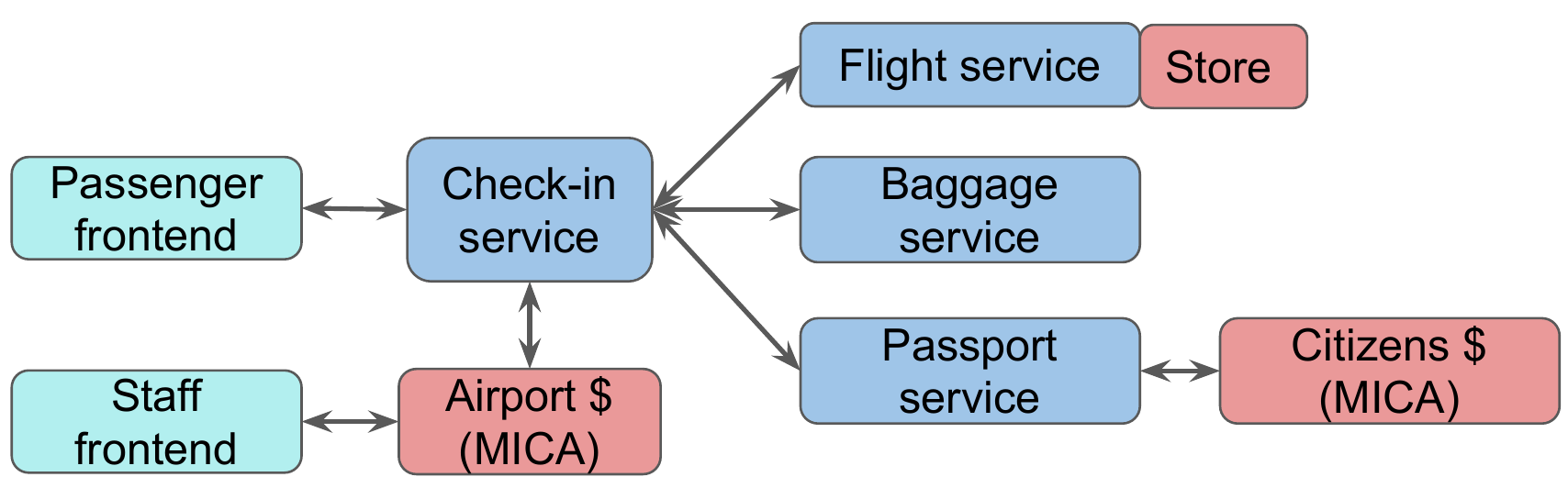}
  	\caption{Flight Registration microservice architecture -- \normalfont{the passenger front-end generates uniformly random passenger registration requests to the Check-in service. The Check-in service then consults the Flight service for flight information data, the Baggage service for the status of the passenger's baggage, and the Passport service to check the passenger's identity. The Passport service issues nested requests to the Citizens database (based on MICA~\cite{MICA}). Upon receiving all responses, the Check-in service registers the passenger in the Airport database (also based on MICA cache). The latter is additionally accessible by the Staff front-end, which is used to asynchronously check all records in the system.}}
    \label{fig:eval_5}
\end{figure}

Finally, we evaluate Dagger on an end-to-end application built with microservices. We design a simple multi-tier service with 8 microservices which implements a Flight Registration service, shown in Figure~\ref{fig:eval_5}.

We design the Flight Registration service such that it exhibits different types of dependencies across tiers (one-to-one, one-to-many, many-to-one), and includes both chain and fanout dependencies. All services communicate with each other over RPC calls and run different threading models to show the flexibility of Dagger. In particular, the \textit{Passenger} and \textit{Staff Frontend} services run non-blocking RPCs to avoid throughput bottlenecks due to high request propagation times. Similarly, the \textit{Check-in} service issues non-blocking requests to the \textit{Flight}, \textit{Baggage}, and \textit{Passport} services, but it later blocks until it receives all the responses before proceeding with blocking calls to the \textit{Airport} service. The \textit{Passport} service also runs blocking RPCs to the Citizens database. In our first experiment, each RPC server processes requests in the dispatch thread to benefit from its low-latency, zero-copy operation.

\begin{figure}[h]
    \centering
    \setlength{\belowcaptionskip}{-10pt}
    \includegraphics[width=1\linewidth]{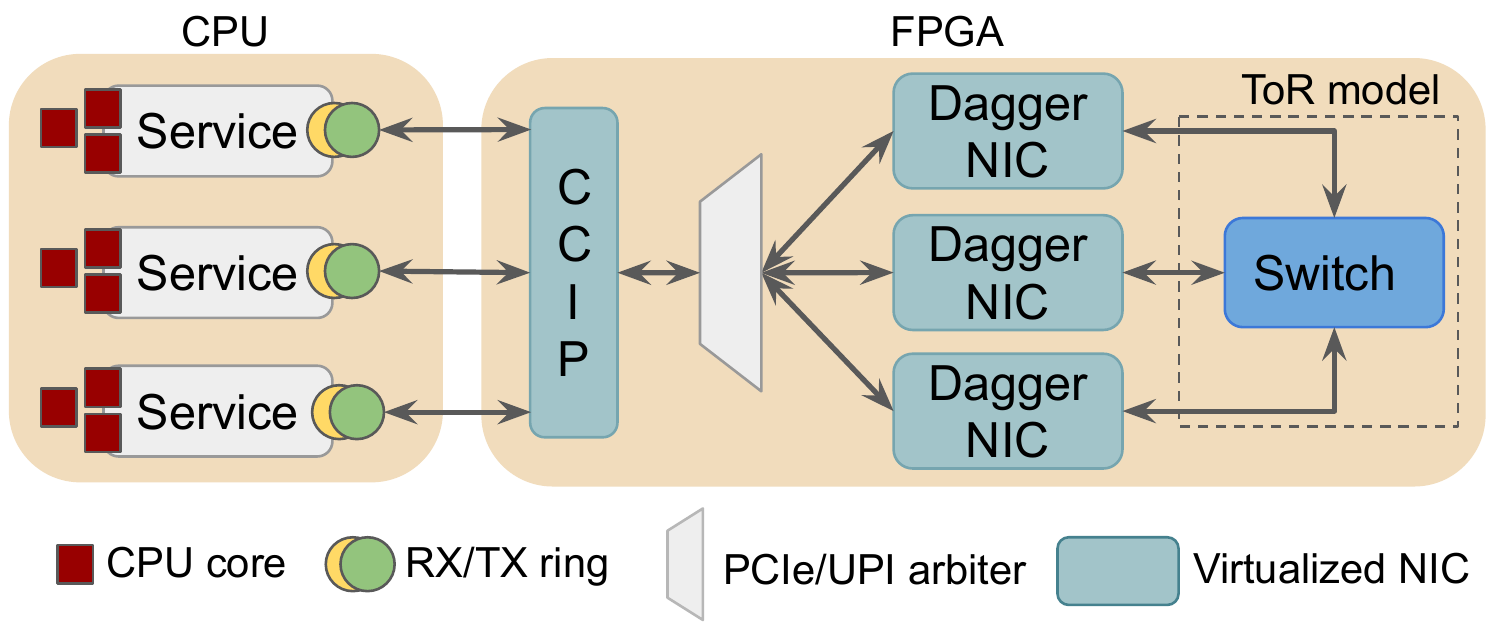}
  	\caption{Virtualization of the Dagger NIC to serve multiple tiers running on the same physical machine -- \normalfont{the PCIe/UPI arbiter provides fair round-robin sharing of the CCI-P bus between tenants; the Switch performs simple L2 packet switching based on the pre-defined static switching table.}}
    \label{fig:virtualization}
\end{figure}

Similarly to our previous experiments, and due to the limitations of the cluster, we run all services on the same machine and over the same physical FPGA. Such a setup requires virtualization pf the NIC to ensure that each tier gets an independent Dagger NIC therefore reflecting the real distributed setup when each tier runs on a separate physical or virtual machine. By placing all tiers on the same machine we additionally show how the Dagger NIC can be virtualized. Our NIC is virtualizable by putting multiple instances of it on the same FPGA and giving them fair round-robin sharing of the system bus and memory. Each instance of the NIC is serving a dedicated microservice tier, i.e., in this experiment, we instantiate 8 copies of the Dagger NIC, as shown in Figure~\ref{fig:virtualization}. The NICs are connected with each other over our simple model of a ToR networking switch with a static switching table. In Section~\ref{sec:discussions}, we give more details on virtualization and show how different microservice tiers can benefit from it.

As previously mentioned, we first run the Flight Registration service with the \textit{Simple} threading model in which each service handles RPC requests directly in dispatch threads along with the networking I/O. The results of this experiment are summarized in the first row in Table~\ref{tab:microservces_results}.

\begin{table}[h]
\begin{threeparttable}
\caption{Summary of performance results for the Flight Registration service. }
\label{tab:microservces_results}
\begin{tabular}{l|l|l|l|l}
\hline
\multicolumn{1}{c|}{\multirow{2}{*}{\begin{tabular}[c]{@{}c@{}}Threading\\ model\end{tabular}}} & \multicolumn{1}{c|}{\multirow{2}{*}{\begin{tabular}[c]{@{}c@{}}Highest load,\\ Krps \tnote{1} \end{tabular}}} & \multicolumn{3}{c}{Lowest latency, us} \\ \cline{3-5}
\multicolumn{1}{c|}{}                                                                           & \multicolumn{1}{c|}{}                                                                              & Median        & 90th       & 99th       \\ \hline \hline
Simple                                                                                           & 2.7                                                                                                & 13.3          & 20.2       & 23.8       \\ \hline
Optimized                                                                                       & 48                                                                                                 & 23.4          & 27.3       & 33.6     \\ \hline
\end{tabular}
\begin{tablenotes}\footnotesize
\item[1] Recorded when the total number of request drops across all tiers does not exceed 1\%. 
\end{tablenotes}
\end{threeparttable}
\end{table}

\begin{figure}[h]
    \centering
    \setlength{\belowcaptionskip}{-5pt}
    \includegraphics[width=1\linewidth]{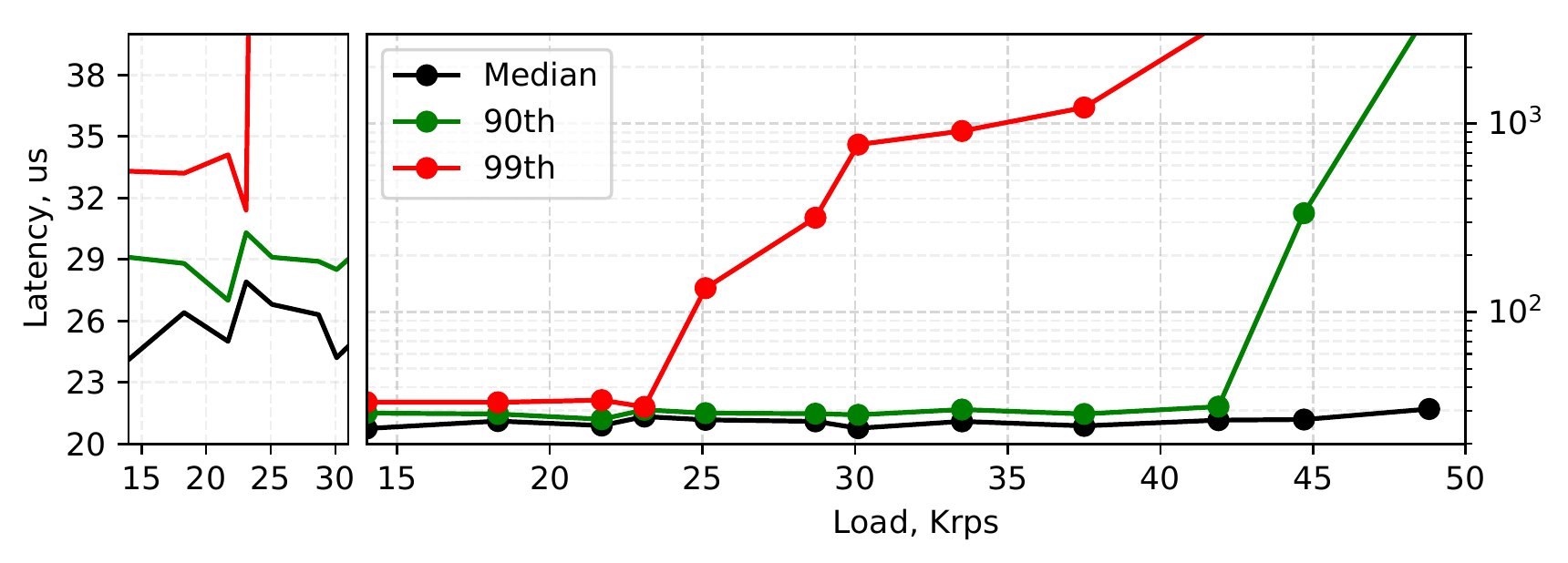}
    \vspace*{-6mm}
  	\caption{Latency/load curves for the Flight Registration service with the \textit{Optimized} threading model -- \normalfont{the figure on the left shows a zoomed-in view of the latency over the load up to 30 Kqps in linear scale. }}
    \label{fig:microservices_eval}
\end{figure}

As seen from Table~\ref{tab:microservces_results}, the maximum throughput with the \textit{Simple} threading model is limited to $2.7 Krps$, however, the system shows low $\mu$s-scale end-to-end latency. In order to profile the application, we design a lightweight request tracing system and integrate it with Dagger. Our analysis reveals that the system is bottlenecked by the resource-demanding and long-running \textit{Flight} service. Handling such RPCs in dispatch threads limits the overall throughput since they block the NIC's RX rings from reading new requests. One well-known way~\cite{RAMCloud} to address this issue is to use different dispatch and worker threads for networking IO and RPC handling. Similar intuition applies to the \textit{Check-in} and \textit{Passport} services. Those are not resource-intensive, but they issue multiple nested blocking RPC calls, and therefore run for a relatively long time. In our next experiment, we configure the \textit{Flight}, \textit{Check-in}, and \textit{Passport} service's RPC servers to run request processing in worker threads (\textit{Optimized} threading model). The results in the second row in Table~\ref{tab:microservces_results} show that such a change in the threading model dramatically increases the overall application throughput by up to 17$\times$. Note that the latency became larger in this case due to the overhead of inter-thread communication and additional request queueing between the dispatch and worker threads. Figure~\ref{fig:microservices_eval} shows a more comprehensive view onto the system behaviour when running with the \textit{Optimized} threading model. The 8-tier Flight Registration service running over Dagger achieves a median and tail latency of $23$ and $33 us$ respectively before the throughput saturation point of $25 Krps$. When throughput crosses the saturation point, the tail latency soars sharply, while the median latency stays at the level of $23 - 26 us$. We do not increase the load further due to the quickly growing number of request drops after that point.

In addition to configuring the software part of Dagger to run different threading models, we also configure the NICs differently. All services in our Flight Registration benchmark besides the \textit{Airport} and \textit{Citizens} services are stateless and they do not cache any data. For this reason, the round-robin dynamic RPC load balancer works best for them and we configure the corresponding NICs to use it. In contrast, the two mentioned services are stateful, and they run the MICA KVS in the backend. However, MICA does not work correctly with round-robin/random load balancers due to the way it partitions the object heap across CPU cores/NIC flows. MICA requires that all requests with the same keys always go to the same partition, and in the original work, it uses Flow Director to steer requests to cores/partitions. In this experiment, we implement our own application-specific Object-Level~\cite{MICA} load balancer for MICA tiers by applying the hash function to each request's key on the FPGA before steering them to the flow FIFOs, and instantiate it inside the Dagger NICs serving the corresponding tiers. This shows how NICs running hardware-offloaded RPCs on FPGAs can be flexibly programmable to satisfy the needs of different applications. More hardware parameters of the NICs can be further fine-tuned for each individual microservice as we briefly discuss in Section~\ref{sec:discussions}.

%% file: Discussions.tex
\section{Discussion}
\label{sec:discussions}

\new{As shown in Dagger's evaluation, relying on memory interconnects for high performance datacenter networking is beneficial, however one might argue that integrating FPGAs/NICs over conventional PCIe busses imposes fewer constraints over the type of CPU a system can use. Additionally, while PCIe is more widely adopted as a peripheral interconnect in today's processors, this trend is increasingly changing. First, the UPI/QPI interconnect is natively supported by all modern datacenter processors (e.g., Xeon family), and any FPGA which implements the UPI/QPI specification can be integrated in it. As of today, we are aware of two such FPGA families: Intel Broadwell Arria 10 (used in this work) and the new Stratix 10 DX device. A similar technology is also being developed by IBM. Their OpenCAPI~\cite{OpenCAPI} cache-coherent CPU-FPGA interface is already used in recent research work on disaggregated memory~\cite{ibm_disaggr}. Second, there is an ongoing collaborative effort from multiple hardware vendors to derive a specification for a new peripheral interface with cache coherency support (CXL) for future devices. Similar efforts in academia have yielded systems like ETH's Enzian~\cite{ETHEnzian}, which closely couples an FPGA with an ARM-based datacenter CPU over the Cavium coherent interconnect CCPI~\cite{cavium}. We believe, Enzian can also be a good physical medium for Dagger. }

\new{Virtualization of network interfaces is another topic around Dagger's design. NIC virtualization enables efficient sharing of a single physical interface between multiple tenants, such as different guest operating systems. Given that Dagger is based on an FPGA and it can be tuned for different applications based on their network characteristics and requirements, it provides an excellent framework to enable virtualization. As seen from Table~\ref{tab:implementation_details}, the Dagger NIC occupies less than 20\% of the available FPGA space when synthesized with a relatively large number of flows and connection cache space. This demonstrates that the same FPGA device can accommodate multiple instances of the NIC at the same time as we show in Section~\ref{sec:eval_microservices}. Each instance can be used as a ``virtual but physical'' NIC for the corresponding tenant, and it can be configured based on the network provisioning requirements of each tenant. }

\new{Additionally, we note that the BRAM memory of FPGAs is one of the key resources enabling reconfigurability and efficiency in Dagger. By leveraging the FPGA to manage on-chip memory one can flexibly split the available memory capacity (53Mb for the FPGA used in Dagger) at very fine granularity, therefore improving the efficiency of NIC caches which is crucial, since NIC cache misses are one of the main performance bottlenecks in commercial NICs~\cite{rdma_1}. This is especially important in the aforementioned virtualized environment. For example, with FPGAs, it is possible to allocate more connection cache memory for NIC instances serving tenants with a large number of connections, or more packet buffer space for tenants experiencing large network footprints. Such on-chip NIC cache management can be easily done at the NIC instance granularity. }

\new{Finally, Dagger-like designs enable efficiently co-designing RPC stacks with transactions in hardware. For instance, in our example of the Flight Registration application in Section~\ref{sec:eval_microservices}, the \textit{Airport} service concurrently processes requests from both the \textit{Check-in} service and \textit{Staff Frontend}. As of now, we implement a lock-based concurrency control mechanism in software which comes with certain overheads in the OS. Alternatively, given the fully programmable nature of FPGAs, one can run synchronization protocols at the RPC level, on the Dagger NIC, such that all requests being received by the service are already serialized. Applications with more complicated transactional semantics (e.g., Paxos, 2PC, etc.) can specifically benefit from this support. }

%% file: Conclusions.tex
\section{Conclusion}

\new{Dagger is an efficient and reconfigurable hardware acceleration platform for RPCs, specifically targeting interactive cloud microservices. In addition to showing the benefits of hardware offload for RPCs to reconfigurable FPGAs, we also demonstrate that using memory interconnects instead of PCIe as the NIC I/O interface offers significant benefits. Most importantly, our work shows that such close coupling of programmable networking devices with processors is already feasible today. Dagger achieves $1.3-3.8\times$ better per-core RPC throughput compared to previous DPDK- and RDMA-based solutions, it provides $\mu$-scale latency, and it can be easily ported to third-party applications, such as memcached and MICA, significantly improving their median and tail latencies. We also show the reconfigurability feature of our proposal by running an example of a multi-tier application and tuning both the software and hardware parts of the stack for each individual microservice to get high end-to-end performance. }